\documentclass[preprint2]{exoplanet}
\usepackage{times}

\newcommand{\refs}{\par\noindent\hangindent=1pc\hangafter=1}
\def\mj{$M_{\rm J}\,$}
\def\rj{$R_{\rm J}\,$}
\def\mic{$\mu$m$\,$}
\def\teff{T$_{\rm{eff}}$}
\def\mstar{M$_{\ast}$}
\def\sgreat{\lower2pt\hbox{$\buildrel {\scriptstyle >}
   \over {\scriptstyle\sim}$}}

\begin{document}

\title{\textbf{\LARGE Giant Planet Atmospheres and Spectra}}

\author {\textbf{\large Adam Burrows}}
\affil{\small\em Princeton University}
\author {\textbf{\large Glenn Orton}}
\affil{\small\em Jet Propulsion Laboratory}

\begin{abstract}
\begin{list}{ } {\rightmargin 1in}
\baselineskip = 11pt
\parindent=1pc
{\small Direct measurements of the spectra of extrasolar giant planets
are the keys to determining their physical and chemical nature.  The goal of theory
is to provide the tools and context with which such data are understood.  It is only
by putting spectral observations through the sieve of theory that the promise
of exoplanet research can be realized.  With the new {\em Spitzer} and HST data of 
transiting ``hot Jupiters," we have now dramatically entered the era of 
remote sensing. We are probing their atmospheric compositions and temperature profiles, 
are constraining their atmospheric dynamics, and are investigating their phase 
light curves.  Soon, many non-transiting exoplanets with wide separations 
(analogs of Jupiter) will be imaged and their light curves and spectra measured. 
In this paper, we present the basic physics, chemistry, and spectroscopy 
necessary to model the current direct detections and to develop the 
more sophisticated theories for both close-in and wide-separation extrasolar giant planets
that will be needed in the years to come as exoplanet research accelerates into its future.
\\~\\~\\~}
 
\end{list}
\end{abstract}

\section{INTRODUCTION}

Our understanding of gas giant planets was informed for many decades 
by remote telescopic observations and in situ measurements of 
Jupiter and Saturn.  These detailed investigations provided
a fine-grained view of their atmospheric compositions, temperatures, 
dynamics, and cloud structures.  However, they left us with a parochial 
view of the range of possible orbits, masses, and compositions that  
has now been shattered by the discovery of extrasolar giant planets (EGPs)
in the hundreds.  We have found gas giants in orbits from $\sim$0.02
AU to many AU, with masses from below Neptune's to $\sim$10 \mj, and around
stars from M to F dwarfs.  The corresponding stellar irradiation fluxes 
at the planet vary by a factor of $\sim$10$^5$, and this variation translates
into variations in atmosphere temperatures from $\sim$100 K to $\sim$2500 K.
With such a range of temperatures and of orbital distances, masses, and ages, 
atmospheres can have starkly different compositions, can be clear or cloudy, 
and can evince dramatic day-night contrasts.  

One must distinguish imaging of the planet itself
by separating the light of planet and star, something that can currently
be contemplated only for wide-separation planets, from measurements of the summed light
when the orbit is tight and the planet can not be separately imaged.
In the latter case, the planet's light can be a non-trivial fraction
of the total, particularly in the infrared.  When transiting, such hot Jupiter systems
provide an unprecedented opportunty to measure the planet's emissions by the
difference in the summed light of planet and star in and out of secondary eclipse
and by the phase variation of that sum.  Generally, the star itself will not vary with the period of the
planet's orbit.  Moreover, in a complementary, but different, fashion, the
wavelength dependence of the transit depth is now being used to probe
the composition of the planet's atmosphere near the terminators. The extrasolar 
giant planets, by dint of their mass and luminosity, have been the first discovered, 
and will serve as stepping stones to the extrasolar terrestrial planets.

To understand in physical detail the growing bestiary of EGPs requires chemistry to 
determine compositions, molecular and atomic spectroscopy to derive opacities, 
radiative transfer to predict spectra, hydrodynamics to constrain atmospheric
dynamics and heat redistribution, and cloud physics.  In short, global 3D radiation-hydrodynamic
general circulation models (GCMs) with multi-spectral, multi-angle, and non-equilibrium
chemistry and kinetics will be needed.  We are not there yet, but basic treatments
have emerged that allow us to interpret and constrain day-night differences, 
profiles, molecular compositions, and phase light curves.  

In this chapter we lay out some of the basic elements 
of any theoretical treatment of the atmospheres, spectra, and light curves of EGPs.  
This theory provides the necessary underpinnings for any progress in EGP studies,
a subject that is engaging an increasing fraction of the world's astronomical and planetary
science communities.  In \S{2.1}, we summarize the techniques for calculating molecular abundances.  We follow in \S{2.2}
with an explication of general methods for assembling opacity tables.  Section {2.3} touches on Rayleigh
scattering, and then we continue in \S{2.4} with a tutorial on albedos and phase functions.
In \S{2.5}, we explain the nature of the transit radius.  Section {2.7} 
contains a very useful analytic model for the atmospheric thermal profile of EGPs, 
which is a generalization for irradiated atmospheres of the classic Milne problem.  
This model incorporates a condition for thermal inversions. Then, in \S{3} we summarize 
lessons learned and knowledge gained from the decades-long study of Jupiter and Saturn.  This includes 
discussions of their spectra, cloud layers, temperatures, and compositions.  Having set the stage, 
we review in \S{4.1} the general chemistry and atmospheric character of EGPs as a function
of orbital distance and age.  This subsection includes a diversion into the putative evolution of Jupiter itself. 
We follow this in \S{4.2} with a few paragraphs on theoretical EGP planet/star flux ratios as a 
function of wavelength and distance, focusing on wide-separation ($>0.2$ AU) EGPs.  Then, in \S{4.3} 
we present highlights from recent campaigns of direct detection of transiting EGP atmospheres, 
with an emphasis on secondary eclipse measurements, the compositions inferred, 
atmospheric temperatures, and thermal inversions.  Finally, in \S{5} we list some 
of the outstanding open issues and future prospects in the study of EGP atmospheres and spectra.

\bigskip
\centerline{\textbf{ 2. MODELING CONCEPTS AND EQUATIONS}}
\bigskip

In this section, we present the core ingredients necessary to construct
theories of the atmospheres and spectra of EGPs.  These includes general chemistry, opacities,
and simple formalisms for the calculation of albedos and phase light curves. We discuss
the concept of ``transit radius" and include an analytic theory for 
the temperature profiles of irradiated exoplanets. The resulting approximate 
equations make clear the key role of opacity and its wavelength dependence 
in determining the character of EGP thermal profiles, in particular in creating 
thermal inversions when they arise.  Such inversions have been inferred for many of the 
hot Jupiters seen in secondary eclipse and are emerging as one of the most exciting and puzzling
features of current exoplanet research.

\bigskip
\noindent
\textbf{ 2.1 Calculation of Atomic and Molecular Abundances in Chemical Equilibrium}
\bigskip

Before any opacities or atmopsheric models can be calculated, the abundances of a mixture of
a large number of species have to be determined for the given temperature
and pressure. The assumption of chemical equilibrium is a good starting point,
though non-equilibrium kinetics may play a role.  Despite this, we present here a straightforward 
discussion of such calculations. Much of this presentation on abundances and that of \S{2.2} on opacities 
can be found in Burrows \& Sharp (1999) or Sharp \& Burrows (2007), to which 
the reader is referred. In addition, there are excellent reviews and/or accounts in Lodders (1999)
Fegley \& Lodders (1994,2001), Lodders \& Fegley (2002), and Sharp \& Huebner (1990). 

For a given temperature, pressure, and composition, the equilibrium abundances of the
various species can be determined by minimizing the total Gibbs free energy
of the system.  This requires a knowledge of the free energy of each species
as a function of temperature, which is normally obtained from thermodynamic
data.  At the temperatures for which data are tabulated, least-square fits can be made for a set of polynomials
whose highest order is given by

\begin{equation}
\Delta G_{pi}(T) = aT^{-1} + b + cT + dT^2 + eT^3\, ,
\label{fitgibbs}
\end{equation}
where $a$, $b$, $c$, $d$, and $e$ are fitted coefficients,
and $\Delta G_{pi}(T)$ is the fitted Gibbs free energy of formation
at temperature $T$ of species $i$ in phase $p$.  The polynomials are
evaluated at the tabulated points and the deviations from the tabulated
values are obtained.

In performing the calculation for a particular temperature, pressure,
and composition, the Gibbs free energy for each species 
is obtained from the database using the fitted coefficients at the
temperature required, then the total free energy of the system is
minimized to obtain the abundances of the gas-phase species, together
with any condensates.  The total dimensionless free energy is given by

\begin{eqnarray}
\frac{G(T)}{RT} = \sum_{i=1}^m \left[n_{1i}
\left\{ \frac{\Delta G_{1i}(T)}{RT} + \ln P +
\ln\left(\frac{n_{1i}}{N}\right)
\right\} \right] + \\ \nonumber
\frac{1}{RT} \sum_{p=2}^{s+1}
\Big[n_{p1} \Delta G_{p1}(T)\Big]\, ,
\label{eqmin}
\end{eqnarray}
where $R$ is the gas constant ($k_BN_A$), and for the first sum for
the gas phase with $p=1$ , $P$ is the total pressure in atmospheres,
$N$ is the number of moles, $m$ is the number of species,
$n_{1i}$ is the number of moles of species $i$, and $\Delta G_{1i}(T)$
is the corresponding free energy of that species.  The second sum is
over the $s$ condensed phases, which may include multiple phases
of the same species, but except at a phase boundary, only one phase
of a particular species in a condensed form is assumed present at any
time, though one may consider solid or liquid solutions.
Consequently, $n_{p1}$ is the number of moles of a condensed
species and $\Delta G_{p1}(T)$ is the free energy of
that species.  Since there is only one species per phase, for
convenience we generally set $i$ equal to 1.

A subset of 30 gas-phase species out of nearly 350 gas-phase
species are usually the most important and selected for detailed treatment.
The species are the neutral atoms: H, He, Li, Na, K, Rb,
Cs, Al, Ca, and Fe, the ions: e$^-$, H$^+$, and H$^-$,
the metal hydrides: MgH, CaH, FeH, CrH and TiH, with the
remaining molecules being H$_2$, N$_2$, CO, SiO, TiO, VO,
CaOH, H$_2$O, H$_2$S, NH$_3$, PH$_3$, and CH$_4$.

Figure \ref{f17.sharp} depicts a representative result (here at 1 atmosphere
pressure and solar elemental abundances) for the temperature dependence of
the equilibrium mixing ratios of water, methane, carbon monoxide, hydrogen
sulfide, phosphene, molecular nitrogen, ammonia, TiO, and VO.  The last two species
may or may not play a role in EGP atmospheres, whereas the others certainly do.

\bigskip
\noindent
\textbf{ 2.2 Calculation of Atomic and Molecular Opacities}
\bigskip

The calculation of the absorption cross sections and opacities of molecules
is made more difficult than that for atoms by the substantially larger
number of transitions and levels involved.  Polyatomic species can have 
hundreds of millions, even billions, of vibrational and rotational lines, 
multiple electronic states, and a complicating mix of isotopes.  
Since it is not possible to measure with precision many transitions to 
determine their oscillator strengths or Einstein $A$ coefficients, 
ab initio calculations using quantum chemical techniques are frequently necessary.
Such calculations can be, and frequently are, calibrated 
with only a few measurements at selected wavelengths, but 
the experimental determination of the quantum numbers of the upper and lower 
states of even a given measured transition can be ambiguous.
Moreover, particularly for hot Jupiters, the high temperatures experienced
require a knowledge of absorption transitions from excited states, 
the so-called ``hot bands," for which there are rarely measurements.
For instance, the methane hot bands and some of the hot bands of water 
are completely unconstrained by current experiment, despite 
the fact that both water and methane are important greenhouse gases 
in the Earth's atmosphere.  

The situation is made even more difficult by the almost complete absence
of calculations or measurements of line broadening coefficients or line profiles.
The upshot is that theorists rely on imperfect and by-and-large uncalibrated
compilations of theoretical line lists, and very approximate theories for 
line broadening.  Some molecules are done better than others, but even for
atoms, for which theory and measurement are rather better, line shapes are 
not well constrained.  In the context of EGPs, this is particularly 
relevant for the alkali metals.

Despite these drawbacks, a rather sophisticated and extensive database of opacities
for the constituents of EGPs has been assembled.  In this subsection, taken 
in part from the review paper by Sharp \& Burrows (2007), we summarize the techniques
and methodolgies needed to derive these opacities and use cgs units for specificity. 
Another useful review is that of Freedman, Marley, \& Lodders (2008).

The calculation of the line strengths for each line of each species depends on
the data available for the species being considered, so different methods have
to be used. In order to reduce the chances of errors with input data in different forms, it is recommended to
convert, if necessary, all the line strengths into the same uniform system,
with the best being integrated line strengths in cm$^2$s$^{-1}$ species$^{-1}$.
These depend only on the temperature.  The lines should then be broadened
into a profile which is dependent on the pressure, then the
absorption in cm$^2$species$^{-1}$ across the profile
should be computed, summing the contributions from any overlapping 
profiles.  The absorption for each species obtained in this manner
depends on the temperature and pressure.  The total opacity
of the gas is obtained by summing the individual contributions weighted
by the corresponding number densities (in cm$^{-3}$) for each of the species,
yielding the total volume opacity in cm$^2$cm$^{-3}$, i.e.
cm$^{-1}$. However, the total mass opacity in cm$^2$g$^{-1}$
is usually the required result, and is obtained by dividing the volume
opacity by the gas mass density.

In its most general LTE (Local Thermodynamic Equilibrium) form,
the integrated strength $S$ of a spectral line in
cm$^2$ s$^{-1}$ species$^{-1}$ is

\begin{equation}
S = \frac{\pi e^2 g_if_{ij}}{m_ec} \frac{e^{-hcF_i/kT}}{Q(T)}
\left[1 - e^{-hc(F_j-F_i)/kT}\right]\, ,
\label{oscil}
\end{equation}
where $g_i$ and $f_{ij}$ are the statistical weight of
the $i^{th}$ energy level and the oscillator strength for a transition from
that level to a higher level $j$, $F_i$, and $F_j$ are the term
values (excitation energies) in cm$^{-1}$ of the $i^{th}$ and $j^{th}$ levels
participating in the transition, and $Q(T)$ is the partition function of the
species at some temperature $T$.  The other symbols have their usual
meanings.  Note that the first term in eq. (\ref{oscil}) gives the line
strength in cm$^2$ s$^{-1}$ absorber$^{-1}$, the next term with the
Boltzmann factor and the partition function converts this to the required
line strength, and the last term is the stimulated emission correction factor,
where $F_j-F_i$ is the transition frequency in wavenumbers, i.e.
$\bar{\nu}$ in cm$^{-1}$.  Although monochromatic opacities are frequently displayed
as functions of wavelength, it is recommended that all opacity
calculations be performed internally in wavenumbers, even if some
of the input data are given in wavelengths, since most molecular spectroscopic
constants and energy levels are expressed in cm$^{-1}$, and adopting a
uniform system of units reduces the chances of error.
Note that some of the data available are not expressed in the form of
oscillator strengths and statistical weights, collectively given as
$gf$-values, but in other forms that must be converted to the required
lines strengths.

The general expression for calculating the partition function is given by

\begin{equation}
Q(T) = \sum_{i=1}^n g_ie^{-hcF_i/kT}\, ,
\label{partfun}
\end{equation}
where the summation is performed over the first $n$ levels, whose
contributions are required at the highest temperatures of interest.  The
term value of the lowest level $F_1$, i.e. the gound state, is zero by
construction.

Figure \ref{fig1.sharp} portrays a representative set of absorption cross sections 
as a function of wavelength in the infrared for the vibration-rotation 
transitions of H$_2$O, NH$_3$, and CH$_4$. Many of the most important absorption bands
in EGP atmospheres are shown and these were calculated using the formalism described 
in this subsection. 

\bigskip
\noindent
\textbf{ 2.3 Rayleigh Scattering}
\bigskip

Rayleigh scattering is a conservative scattering process by atoms
and molecules.  Although strong in the ultraviolet/blue, the scattering
cross sections quickly fade toward the red region of the spectrum
($\propto \lambda^{-4}$).  Rayleigh scattering has little effect on the
spectra of isolated brown dwarfs, but irradiated EGPs reflect a
non-zero fraction of the incident intensity.

The Rayleigh scattering cross sections are derived from polarizabilities,
which are in turn derived from refractive indices.
The refractive indices are readily available at 5893 \AA\ (Na D)
and are assumed not to vary strongly with wavelength.
The Rayleigh cross sections are derived via,
\begin{equation}
\sigma_{Ray} = {8\over 3}\pi k^4 \left({{n-1}\over{2\pi L_0}}
\right)^2\, ,
\label{rayleigh}
\end{equation}
where $n$ is the index of refraction, $k$ is the wavenumber
(2$\pi/\lambda$), and $L_0$ is Loschmidt's number, the number of molecules
per cubic centimeter at STP (= 2.687$\times$10$^{19}$).
Given the strong inverse dependence on wavelength of eq. (\ref{rayleigh}),
Rayleigh scattering is most pronounced in the blue and ultraviolet and 
is ultimately responsible for the Earth's blue sky.

\bigskip
\noindent
\textbf{ 2.4 Albedos and Phase Functions}
\bigskip

Objects in the solar system, such as planets, asteroids, and moons, are
seen and studied in reflected solar light.  The brightness of the reflection
depends upon the orbital distance, the stellar flux, the reflectivity of the 
object, the detector angle (the ``phase angle"), and the object's radius.
The reflectivity, in the guise of an ``albedo" (defined below), 
bears the stamp of the composition of its surface and/or atmosphere, and its wavelength
dependence is a distinctive and discriminating signature.  
Solar-system objects are too cold to emit much in the optical, where the Sun
is brightest, but emit in the mid-infrared, where the peak of a $\sim$40-800 K
black body resides.  Hence, there is a simple and obvious separation in their
spectra between reflection and emission bumps that allows an unambiguous 
definition of the albedo and its interpretation as a dimensionless 
reflectivity bounded by a value of ``one" (however, see below), the latter implying full 
reflection and no absorption.  

However, some EGPs, the close-in and transiting
variants, are so near their central stars that their surface and atmospheric 
temperatures can be quite large ($\sim$1000-2500 K). Hot planets (due 
to either proximity or youth) can be self-luminous in the near infrared (and even 
in the optical).  As a result, the reflection and emission components
can overlap in wavelength space, and are not so cleanly separated, as they are 
for solar-system objects.  The upshot is that the albedo and ``reflectivity" 
might be misnomers, particularly in the near-IR. Nevertheless, 
the planet/star flux ratio as a function of wavelength is an important 
probe of EGP atmospheres and has to date been used with profit to diagnose 
their thermal and compositional character. This is true despite the fact
that the associated albedos could be far above one at some wavelengths\footnote{Note
that if the complete radiative transfer solution with stellar irradiation
is derived, the concept of an albedo is redundant and unnecessary.}.  With this caveat 
in mind and the traditional interpretation intact for planets at greater 
orbital distances and older ages similar to the Jupiter/Sun pair,
we proceed to develop the formalism by which the planet/star flux ratio
is calculated, the albedo is defined, and why they are important.  In the process we 
distinguish the geometric albedo ($A_g$), the spherical albedo ($A_s$), and the Bond
albedo ($A_B$), and connect them to the planet/star flux ratio, $F_p/F_*$. 
We also introduce the ``phase function," by which the observed light 
curve is described.  Much of the development below is taken from 
papers by Sudarsky, Burrows, \& Pinto (2000), Sudarsky, Burrows, Hubeny, \& Li (2005),
and Burrows, Sudarsky, \& Hubeny (2004).  The papers by Marley et al. (1999) 
and Burrows, Ibgui, \& Hubeny (2008) are also good resources on this general topic.

Planetary phase is a function of the observer-planet-star orientation,
and the angle whose vertex lies at the planet is known as the {\it phase
angle} ($\alpha$).  The formalism for the computation of
planetary brightness as a function of phase angle has been presented
by numerous authors.  Following Sobolev (1975), one can relate the
planetary latitude ($\psi$) and longitude ($\xi$)
to the cosine of the angle of incident radiation ($\mu_0$) and the cosine
of the angle of emergent radiation ($\mu$) at each point on the planet's
surface:
\begin{equation}
\mu_0 = \cos\psi\cos(\alpha-\xi)
\label{eq_latlong}
\end{equation}
and
\begin{equation}
\mu = \cos\psi\cos\xi\, ,
\end{equation}
where latitude is measured from the orbital plane and longitude is measured
from the observer's line of sight.  The phase angle is then,
\begin{equation}
\alpha = \cos^{-1}\left(\mu\mu_0 - \left[{(1-\mu^2)(1-\mu_0^2)}\right]^{1/2}cos\phi\right)\, ,
\end{equation}
where $\phi$ is the azimuthal angle between the incident and
emergent radiation at a point on the planet's surface.  The emergent
intensity from a given planetary latitude and longitude
is 
\begin{equation}
I(\mu,\mu_0,\phi) = \mu_0S\rho(\mu,\mu_0,\phi),
\end{equation}
where the incident flux on a patch of the planet's surface
is $\pi\mu_0S$, and $\rho(\mu,\mu_0,\phi)$ is the reflection coefficient.
In order to compute the energy reflected off the entire planet, one must
integrate over the surface of the planet.  For a given planetary phase,
the energy per time per unit area per unit solid angle received by an observer is
\begin{eqnarray}
E(\alpha) = 2S{R_p^2\over{d^2}}\int_{\alpha-\pi/2}^{\pi/2}\cos(\alpha-\xi)
\cos(\xi)d\xi \times \\ \nonumber
\int_0^{\pi/2}\rho(\mu,\mu_0,\phi)\cos^3\psi
d\psi \, , 
\end{eqnarray}
where $R_p$ is the planet's radius and $d$ is the distance to the
observer.  This quantity is related to the {\it geometric albedo} ($A_g$), the reflectivity
of an object at full phase ($\alpha=0$) relative to that of a perfect 
Lambert disk (for which $\rho(\mu,\mu_0,\phi) = 1$) of
the same radius under the same incident flux, by
\begin{equation}
A_g = {{E(0)d^2}\over{\pi SR^2}}\,  ,
\end{equation}
where $E(0)$ is $E(\alpha)$ at $\alpha = 0$.
A planet in orbit about its central star displays
a range of phases, and the planet/star flux ratio is given by
\begin{equation}
{F_p\over{F_*}} = A_g\left({R_p\over{a}}\right)^2\Phi(\alpha),
\label{falpha}
\end{equation}
where $\Phi(\alpha)$ is the classical {\it phase function}, which is equal to $E(\alpha)/E(0)$, $R_p$
is the planet's radius, and $a$ is its orbital distance.  
This formula is one of the core relationships in the study of irradiated and reflecting EGPs.

$\Phi(\alpha)$ is normalized to
be 1.0 at full face, thereby defining the geometric albedo,
and is a decreasing function of $\alpha$.  For Lambert
reflection,  an incident ray on a planetary patch emerges uniformly
over the exit hemisphere, $A_g$ is 2/3 for purely scattering atmospheres, and $\Phi(\alpha)$ is given by the
formula:
\begin{equation}
\Phi(\alpha) = \frac{\sin(\alpha) + (\pi - \alpha)\cos(\alpha)}{\pi}\, .
\label{lambert}
\end{equation}
However, EGP atmospheres are absorbing and the anisotropy of
the single scattering phase function for grains, droplets, or molecules
results in non-Lambertian behavior.  For instance, back-scattering off cloud particles
can introduce an ``opposition" effect for which the planet appears anomalously
bright at small $\alpha$. Figure \ref{fig_phaseall} provides some theoretical EGP 
phase functions taken from Sudarsky et al. (2005) in which this effect is clearly seen.  
Figure \ref{realphase} depicts the corresponding phase curves for some the solar system objects. 
These two figures together suggest a likely range for exoplanets.

Both $A_g$ and $\Phi(\alpha)$ are functions of wavelength, but the wavelength-dependence
of $A_g$ is the most extreme.  In fact, for cloud-free atmospheres, due
to strong absorption by molecular bands, $A_g$ can
be as low as 0.03, making such objects very ``black."  Rayleigh 
scattering serves to support $A_g$, but mostly in the blue and
UV, where various exotic trace molecules can decrease it.
The presence of clouds increases $A_g$ significantly.  For instance, at 0.48 \mic, Jupiter's
geometric albedo is $\sim$0.46 and Saturn's is 0.39 (Karkoschka 1999).  However, for orbital 
distances less than 1.5 AU, we expect the atmospheres of most EGPs to be 
clear.  The albedo is correspondingly low.  As a consequence, the
theoretical albedo is very non-monotonic with distance, ranging in the visible 
from perhaps $\sim$0.3 at 0.05 AU, to $\sim$0.05 at 0.2 AU, to $\sim$0.4 at 4 AU, 
to $\sim$0.7 at 15 AU.  In the visible ($\sim$0.55 \mic), the geometric albedo for 
a hot Jupiter is severely suppressed by Na-D at 0.589 \mic.  Due to a methane
feature, the geometric albedo can vary from 0.05 at $\sim$0.6 \mic to $\sim$0.4
at 0.625 \mic.  Hence, variations with wavelength and with orbital
distance by factors of 2 to 10 are expected.

$\Phi(\alpha)$ and $A_g$ must be calculated or measured, but the sole dependence
of $\Phi(\alpha)$ on $\alpha$ belies the complications introduced by an orbit's
inclination angle ($i$), eccentricity ($e$), and argument of periastron ($\omega$).
Along with the period ($P$) and an arbitrary zero of time, these are most of the so-called Keplerian elements of an orbit.
Figures 5 and 6 in Chapter 2 of this volume define these orientational and orbital parameters.
In the plane of the orbit, the angle between the planet and the periastron/periapse (point
of closest approach to the star) at the star is $\theta$. In 
celestial mechanics, $\theta$ is the so-called ``true anomaly."  For an 
edge-on orbit ($i = 90^{\circ}$), and one for which the line of nodes is 
perpendicular to the line of sight (longitude of the ascending node, $\Omega$, equals $90^{\circ}$)
and parallel to the star-periapse line ($\omega = 0^{\circ}$),
$\theta$ is complementary to $\alpha$ ($\alpha = 90^{\circ} - \theta$).  As a result, $\theta = 0^{\circ}$ at
$\alpha = 90^{\circ}$ (greatest elongation) and increases with time.  Also, for
such an edge-on orbit, $\alpha = 0^{\circ}$ at superior conjunction.
In general,
\begin{equation}
\cos(\alpha) = \sin(\theta + \omega)\sin(i)\sin(\Omega) - \cos(\Omega)\cos(\theta + \omega) \, .
\label{cosinef}
\end{equation}
Since it is common to define the observing coordinate system
such that $\Omega = 90^{\circ}$, we have the simpler formula:
\begin{equation}
\label{eq_elements}
\cos(\alpha) = \sin(\theta+\omega)\sin(i).
\end{equation}

In order to produce a model light curve for a planet orbiting its central star,
one must relate the planet's orbital angle ($\theta$, the true anomaly),
as measured from periapse, to the time ($t$) in the planet's
orbit:
\begin{eqnarray}
\label{eq_time}
t(\theta) = {{-(1-e^2)^{1/2}P}\over{2\pi}} 
\Bigl({{e\sin\theta}\over {1+e\cos\theta}}  - \\ \nonumber
2(1-e^2)^{-1/2}\tan^{-1}\left[{{(1-e^2)^{1/2}\tan(\theta/2)} \over{1+e}}\right]\Bigr),
\end{eqnarray}
where $P$ is the orbital period and $e$ is the eccentricity.
By combining eq. (\ref{eq_time}) and (\ref{eq_elements}), we derive the exact
phase of any orbit at any time.

For a circular orbit, $R$ is equal to the semi-major axis ($a$).  However, a planet 
in an eccentric orbit can experience significant variation in $R$, and, therefore,
stellar irradiation (by a factor of $(\frac{1+e}{1-e})^2$).  For example,
if $e = 0.3$, the stellar flux varies by $\sim$3.5 along
its orbit.  For $e$ = 0.6, this variation
is a factor of 16!  Such eccentricities are by no means rare in the sample of known EGPs.
Therefore, in response to a changing stellar flux it is possible 
for the composition of an EGP atmosphere to change significantly
during its orbit, for clouds to appear and disappear, and for there to be lags 
in the accommodation of a planet's atmosphere to a varying irradiation regime.  Ignoring the latter,
eqs. (\ref{falpha}) and (\ref{cosinef}) can be combined with $\Phi(\alpha)$ and the standard Keplerian formula
connecting $\theta$ and time for an orbit with a given $P$ and $e$ to derive an EGP's
light curve as a function of wavelength, $i$, $e$, $\Omega$, $\omega$, and time.
Depending upon orientation and eccentricity, the brightness of an EGP can vary
in its orbit not at all (for a face-on EGP in a circular orbit) or quite dramatically
(e.g., for highly eccentric orbits at high inclination angles).
Since astrometric measurements of stellar wobble
induced by EGPs can yield the entire orbit (including inclination), data from
the Space Interferometry Mission (SIM) (Unwin \& Shao 2000)
or Gaia (Perryman 2003) could provide important
supplementary data to aid in the interpretation of direct detections of EGPs.

The {\it spherical albedo} is the fraction of incident light reflected by a sphere
at all angles.  For a theoretical object with no absorptive opacity, all incident
radiation is scattered, resulting in a spherical albedo of unity.  The
spherical albedo is related to the geometric albedo by
$A_s=qA_g$, where $q$ is the {\it phase integral}:
\begin{equation}
q = 2\int_0^{\pi}\Phi(\alpha)\sin\alpha d\alpha\, .
\label{eq_psratio}
\end{equation}
For isotropic surface reflection (Lambert reflection) $q = \frac 3 2$, while
for pure Rayleigh scattering $q = \frac 4 3$.
Although not written explicitly, all of the above quantities
are functions of wavelength.

Van de Hulst (1974) derived
a solution for the spherical albedo of a planet covered with a semi-infinite
homogeneous cloud layer.  Given a single-scattering albedo of $\sigma$
(= $\sigma_{scat}/\sigma_{total}$) and a scattering asymmetry factor of $g
= <\cos\theta>$ (the average cosine of the scattering angle), van de
Hulst's expression for the spherical albedo of such an atmosphere is 
\begin{equation}
A_s \approx {(1 - 0.139s)(1 - s)\over {1 + 1.170s}}\, ,
\label{vandeh}
\end{equation}
where
\begin{equation}
s = \left[{1 - \sigma}\over {1 - \sigma g}\right]^{1/2}\, .
\end{equation}

The Henyey-Greenstein single-scattering phase function,
\begin{equation}
p(\theta) = {{1-g^2}\over {(1+g^2-2g\cos\theta)^{3/2}}}\, ,
\end{equation}
is frequently used as a fit to the overall scattering phase function,
where again $g = <\cos\theta>$.   Other phase functions can be used,
but their specific angular dependence has been found to be less important than the
value of the integral, $g$, itself.  In addition, Rayleigh and cloud particle 
scattering are both likely to result in significant polarization of the 
reflected light from an EGP (Seager, Whitney, \& Sasselov 2000).  The degree of polarization will depend 
strongly on orbital phase angle and wavelength, and can reach many tens of percent,  However,
polarization may be difficult to measure. To date, there is no credible evidence 
for polarized light from any EGP.  The degree of polarization in the 
optical and UV is expected to be largest.

The important {\it Bond\/} albedo, $A_B$, is the ratio of the total
reflected and total incident powers.  It is obtained by weighting
the spherical albedo by the spectrum of the illuminating source and
integrating over all wavelengths:
\begin{equation}
A_B = {\int_0^\infty A_{s,\lambda} I_{inc,\lambda} d\lambda
\over {\int_0^\infty I_{inc,\lambda} d\lambda}}\, ,
\label{bondeq}
\end{equation}
where the $\lambda$ subscript signifies that the incident intensity varies
with wavelength.

With the Bond albedo, one can make a crude estimate of the ``effective temperature"
of the planet's emission component in response to irradiation.  Under the assumption
that the incident total power is equal to the emitted power, one derives:

\begin{equation}
T_{\rm eff} = T_{*} \left(\frac{fR_{*}}{a}\right)^{1/2}\left(1-A_B\right)^{1/4}\, ,
\label{bondeff}
\end {equation}
where $T_*$ is the stellar effective temperature, $a$ is the orbital distance, $R_*$ is the stellar
radius, and $f$ is a measure of the degree of heat redistribution around the planet.  It is equal
to $1/4$ if the reradiation is isotropic, and $1/2$ if the planet reemits only on the 
day side and uniformly.  Note that $T_{\rm eff}$ is independent of the planet's radius.

Eq. (\ref{bondeff}) is useful in estimating the temperatures achieved
by radiating planets, but has conceptual limitations.  First, it presumes that the planet
is not self-luminous and radiating the residual heat of formation in its core.
This assumption is not true for young and/or massive EGPs. It is not true of Jupiter.  Second, it encourages the 
notion that the reflection and emission peaks are well-separated.  This is not correct for 
the hot Jupiters, for which the two components overlap and merge.  Third, in radiative
transfer theory and equilibrium, the true flux from a planet is the {\it net} flux.  Without 
internal heat sources, this is {\it zero} for irradiated planets.  Finally, eq. (\ref{bondeff})
is often used to determine the temperature of an atmosphere.  However, EGP atmospheres
have temperature {\it profiles}.  In their radiative zones the temperatures can vary
by factors of $\sim$3, as can the effective photospheric temperatures in the near- and mid-IR.
Hence, one should employ eq. (\ref{bondeff}) to obtain atmospheric temperatures 
only when very approximate numbers are desired. Importantly, calculating
the irradiated planet's spectrum with a $T_{\rm eff}$ derived using eq. (\ref{bondeff})
results in large errors across the entire wavelength range that can severely compromise
the interpretation of data.  

\bigskip
\noindent
\textbf{ 2.5 The Transit Radius}
\bigskip

A transiting planet reveals its radius ($R_p$) by the magnitude of the diminution
in the stellar light during the planet's traverse of the stellar disk.  This 
is the primary eclipse.  In fact, it is the ratio of the planet 
and star radii that is most directly measured, so a knowledge of the star's 
radius is central to extracting this important quantity.  With a radius 
and a mass (for transiting planets, the inclination must be near 
90$^{\circ}$ and the inclination degeneracy is broken), we can 
compare with theories of the planet's physical structure and evolution.  
However, since the effective edge of the planetary disk is determined
by the opacity of the atmosphere at the wavelength of observation, the radius 
of a gas giant is wavelength-dependent.  Importantly, the variation with wavelength
of the measured radius can serve as an ersatz atmospheric ``spectrum."  From 
this spectrum, one can determine the atmosphere's constituent atoms and molecules. 
Specifically, the apparent radius is larger at wavelengths for which the opacity
is larger and smaller at wavelengths for which the opacity is smaller.  For example,
the planet should be larger in the sodium D line at $\sim$5890 \AA (as in fact was found for HD 209245b)
than just outside it.  It should be larger near the absorption peaks of the water
spectrum for those planets with atmospheric water than in the corresponding troughs.

However, the ``transit radius" is not the same as the classical photospheric
radius of an atmosphere.  The latter is determined by the depth in the atmosphere 
where $\tau_{\nu} = 2/3$ in the {\em radial} direction.  The transit radius is where
this same condition obtains along the {\em chord} from the star, perpendicular to
the radius to the center.  At the spherical radius where a light beam experiences
$\tau_{\nu} = 2/3$, $\tau_{\nu}$ along the chord can be much larger.  Therefore, 
to achieve the $\tau_{\nu} = 2/3$ condition along the chord pushes the transit radius 
(also referred to as an ``impact parameter") to larger values. It is only 
at this greater altitude and lower pressure that the chord optical depth is $\sim$$2/3$.
An important difference between the transit radius spectrum and the actual ``emission" 
spectrum from a classical photosphere (such as is relevant at secondary eclipse)
is that in the latter case, if the atmosphere were isothermal the spectrum would be a black body
and there would be no composition information.  However, the transit radius spectrum always manifests
the wavelength dependence of the opacities of its atmospheric constituents, even if the
atmosphere were isothermal.  This difference can be exploited to maximize the scientific
return from the study of a given transiting EGP.  In sum, if the atmosphere is extended 
and if the monochromatic opacity at the measurement wavelength is large, the transit radius 
and the photosphere radius can be rather different and the distinction should always be kept in mind.   

A difficulty in interpreting transit radius spectra is that one is probing the 
planetary limb, the terminator.  This means that when comparing with data 
models must incorporate profiles on both the day side and night side, 
and at both the equator and poles. In particular the day side and the 
night side can be at different temperatures and have different compositions.  
This complication is not always appreciated.

One can estimate the magnitude of the excess of the transit radius over the 
photospheric radius using a simple exponential atmosphere (see also Fortney 2005). The 
wavelength-dependent optical depth, $\tau_{\rm chord}$, along a chord
followed by the stellar beam through the planet's upper atmosphere,  is approximately:

\begin{equation}
\tau_{\rm chord} \sim \kappa\rho_{ph} H \sqrt{\frac{2\pi R_p}{H}}\ e^{-(\frac{\Delta R_{ch}}{H})} \, ,
\label{tau_eq}
\end{equation}
where $\kappa$ is the wavelength-dependent opacity,
$\rho_{ph}$ is the mass density at the photosphere, $\Delta R_{ch}$ is the excess radius over
and above the $\tau_{ph} = \frac{2}{3}$ radius (the radius of the
traditional photosphere), and $H$ is the atmospheric density scale height.
The latter is given approximately by $k T/\mu g m_p$, where $\mu$ is the mean molecular weight, $g$
is the surface gravity, $T$ is some representative atmospheric temperature, and $m_p$ is the proton mass.
By definition, and assuming an exponential atmosphere, $\tau_{ph}$ = $\kappa\rho_{ph} H$ = $\frac{2}{3}$.
For $\tau_{\rm chord}$ to equal $\frac{2}{3}$, this yields

\begin{equation}
\Delta R_{ch} = H \ln\sqrt{{\frac{2\pi R_p}{H}}}\\
 \sim 5\ H  \, .
\label{tau_equ}
\end{equation}
This excess can be from $\sim$1\% to $\sim$10\%, depending upon the wavelength, temperature, 
gravity, and deviations from a strictly exponential profile.  It is smallest
for high-gravity planets at larger orbital distances, whereas in the UV near 
Lyman-$\alpha$, and with a planetary wind, this excess for HD 209458b is measured to be a 
factor of 2$-$3.

At times, the transit radius spectrum is called the ``transmission spectrum." 
Since what is measured is the effective area of the planet, not the spectrum 
of light transmitted through the planetary limb, this is a slight misnomer.  
One is measuring the atmospheric edge position (relative to some zero-point) as a function
of wavelength, and not the light transmitted through the finite extent of the atmosphere.  To properly do 
the latter would require a resolved image of the planet that distinguishes the atmosphere from the opaque
central planetary disk. Note also that when the atmospheric opacity is low (and, hence, light is more easily
``transmitted"), the dip in the stellar flux (what is actually measured) is smaller, not larger. Nevertheless,
this rather pedantic point does not inhibit one from profiting from ``transmission spectrum" measurements.

\bigskip
\noindent
\textbf{ 2.6 Analytic Model for the Temperature Profile of an Irradiated Planet}
\bigskip

One can derive a model for the temperature profile 
of an irradiated EGP atmosphere that is a generalization of the classical
Milne problem for an isolated atmosphere.  This model can incorporate
the difference between the opacity to the insolating (``optical") and emitting (``infrared") radiation streams,
contains a theory for stratospheres and thermal inversions, and a condition for 
their emergence. The mathematical development of this analytic theory is taken from 
Hubeny, Burrows, \& Sudarsky (2003), to which the reader is referred
for further details.  The paper by Chevallier, Pelkowski, \& Rutily (2007)
is also of some considerable utility.

The equation of hydrostatic equilibrium is fundamental in atmospheric theory,
and obtains as long as the Mach number of the gas is low.  It can be written:

\begin{eqnarray}
\frac{dP}{dz} = -g\rho \\ \nonumber
\frac{dP}{dm} = g \, ,
\label{hydro}
\end{eqnarray}
where $P$ is the pressure, $g$ is the acceleration due to gravity (generally assumed constant),
$\rho$ is the mass density, $z$ is the altitude, and $m$ is the areal (column) mass density defined
by $dm = -\rho dz$.  Note that for a constant $g$, the pressure and the column mass are directly 
proportional.

If the atmosphere is convective, the temperature gradient follows an adiabat 
and is given by: 

\begin{equation}
\frac{d \ln T}{d \ln P} = C_p/R \, ,
\label{lapse}
\end{equation}
where $C_p$ is the specific heat at constant pressure and $R$ is the gas constant.  Given the temperature and pressure at any
point in the convective region, together with the specific heat (which can be
determined using the composition), the temperature at any atmospheric pressure
can be determined.  With that knowledge, eq. (\ref{lapse}) can then be used to determine the
altitude scale.  Hence, the combination of the equation of hydrostatic 
equilibrium and eq. (\ref{lapse}) can be used to determine 
the adiabatic lapse rate for a given specific heat, pressure, and temperature.

However, for most of a strongly irradiated atmosphere, and in 
radiative zones in general, radiation carries the energy flux.
The radiative transfer equation is written as
\begin{equation}
\label{rte1}
\mu \, \frac{dI_{\nu\mu}}{dm}=\chi_\nu\left(I_{\nu\mu}-S_\nu\right)\, ,
\end{equation}
where $I_{\nu\mu}$ is the specific intensity of radiation as a function
of frequency, $\nu$, angle $\mu$ (the cosine of the angle of propagation with respect to
the normal to the surface), and the geometrical coordinate,
taken here as the column (areal) mass $m$. The monochromatic optical depth is defined
as $d\tau_\nu = \chi_\nu\, dm$.

$S_\nu$ is the source function, given in LTE by
\begin{equation}
\label{sdef}
S_\nu=\frac{\kappa_\nu}{\chi_\nu}B_\nu+\frac{\sigma_\nu}{\chi_\nu}J_\nu\, .
\end{equation}
Here $\kappa_\nu$ is the true absorption coefficient, $\sigma_\nu$ is the
scattering coefficient, $B_\nu$ is the black body function, $J_\nu$
is the zeroth angular moment of the specific intensity, and
$\chi_\nu = \kappa_\nu + \sigma_\nu$ is the total absorption coefficient.
The zeroth moment is equal to the radiation energy density, divided by $4\pi$.
All coefficients are per unit mass.

The first moment of the transfer equation is written
\begin{equation}
\label{Hmom1}
\frac{dH_\nu}{dm}=\chi_\nu\left(J_\nu-S_\nu\right)\, ,
\end{equation}
which can be rewritten, using equation (\ref{sdef}) as
\begin{equation}
\label{Hmom2}
\frac{dH_\nu}{dm}=\kappa_\nu \left(J_\nu - B_\nu\right)\, .
\end{equation}
$H_\nu$ is the first angular moment of 
the specific intensity and is equal to the radiation flux, divided by $4\pi$.
Integrating over frequency one obtains
\begin{equation}
\label{Hmomint}
\frac{dH}{dm} = \kappa_J J - \kappa_B B\, ,
\end{equation}
where $\kappa_J$ and $\kappa_B$ are the absorption and Planck mean
opacities, respectively, defined by
\begin{equation}
\label{kjdef}
\kappa_J = \frac{\int_0^\infty \kappa_\nu J_\nu d\nu }{\int_0^\infty J_\nu d\nu}\, ,
\end{equation}
and
\begin{equation}
\label{kbdef}
\kappa_B = \frac{\int_0^\infty \kappa_\nu B_\nu d\nu}{\int_0^\infty B_\nu d\nu}\, .
\end{equation}
These two opacities are often assumed to be equal. However, one should 
distinguish them here because the difference between $\kappa_J$ and $\kappa_B$
turns out to be crucial in the case of strongly irradiated atmospheres.

The second moment of the transfer equation is
\begin{equation}
\label{Kmom1}
\frac{dK_\nu}{dm}=\chi_\nu H_\nu\, ,
\end{equation}
and integrating over frequency one obtains
\begin{equation}
\label{Kmomint}
\frac{dK}{dm}=\chi_H H\, ,
\end{equation}
where
\begin{equation}
\chi_H = \frac{\int_0^\infty \chi_\nu H_\nu d\nu}{ \int_0^\infty H_\nu d\nu}
\end{equation}
which is referred to as the flux mean opacity.

Finally, the radiative equilibrium equation is written as
\begin{equation}
\label{re1}
\int_0^\infty \kappa_\nu \left(J_\nu - B_\nu\right) d\nu = 0\, .
\end{equation}
Using the above mean opacities, this can be rewritten
\begin{equation}
\label{reint}
\kappa_J J - \kappa_B B = 0\, .
\end{equation}
Substituting (\ref{reint}) into (\ref{Hmomint}), one obtains another
form of the radiative equilibrium equation,
\begin{equation}
\frac{dH}{dm} = 0\, ,\quad {\rm or} \quad H={\rm const}\equiv (\sigma/4\pi)\,
T_{\rm eff}^4\, ,
\end{equation}
where $\sigma$ is the Stefan-Boltzmann constant.

From equation (\ref{reint}), one has
$B=(\kappa_J/\kappa_B) J$, which yields an expression for $T$ through
$J$ using the well-known relation $B=\sigma T^4$. To determine $J$, one uses
the solution of the second moment of the transfer equation
$K(\tau_H) = H \tau_H = (\sigma/4\pi)\, T_{\rm eff}^4\, \tau_H$,
where $\tau_H$ is the optical depth using the flux-mean opacity,
and expresses the moment $K$ through $J$ by means of the Eddington
factor, $f_K \equiv K/J$. Similarly, one expresses the surface flux through
the second Eddington factor, $f_H \equiv H(0)/J(0)$ (see also Hubeny 1990):
\begin{equation}
T^4 = \frac{3}{4}\, T_{\rm eff}^4\, \frac{\kappa_J}{\kappa_B}
\left[ \frac{1}{3 f_K} \tau_H + \frac{1}{3 f_H} \right] +
\frac{\kappa_J}{\kappa_B}\, W\, T_\ast^4\, .
\end{equation}
where $W$ is the dilution factor, $(R_*/a)^2$.
This solution is exact within LTE and is the generalization of the classical Milne atmosphere solution.

The usual LTE-gray model
consists in assuming all the mean opacities to be equal to the
Rosseland mean opacity. If one adopts the Eddington approximation
($f_K = 1/3; f_H = 1/\sqrt 3$), then one obtains a simple expression
\begin{equation}
T^4 = \frac{3}{4}\, T_{\rm eff}^4
\left(\tau + 1/\sqrt 3 \right) + W T_\ast^4\, .
\end{equation}

We will consider the most interesting case, namely strong irradiation,
defined by $W T_\ast^4 \gg T_{\rm eff}^4$. In this case, the second
term in the brackets is negligible, and one may define a {\em penetration depth} as
the optical depth where the usual thermal part ($\propto T_{\rm eff}^4$)
and the irradiation part ($\propto W T_\ast^4$) are nearly equal, to whit
\begin{equation}
\tau_{\rm pen} = W\, \left(\frac{T_\ast}{T_{\rm eff}}\right)^4\, .
\end{equation}
The behavior of the local temperature in the case of a strict gray model
is very simple -- it is essentially constant, $T= T_0\equiv W^{1/4} T_\ast$ for
$\tau < \tau_{\rm pen}$, and follows the usual distribution
$T \propto \tau^{1/4} T_{\rm eff}$ in deep layers, $\tau > \tau_{\rm pen}$.
In the general case, one has to retain the ratio of the
absorption and Planck mean (assuming still that the flux mean opacity
is well approximated by the Rosseland mean). In the irradiation-dominated
layers ($\tau < \tau_{\rm pen}$), the temperature is given by
\begin{equation}
\label{tupp}
T = \gamma\, W^{1/4}\, T_\ast\, ,
\end{equation}
where
\begin{equation}
\gamma \equiv ({\kappa_J}/{\kappa_B})^{1/4}\, .
\end{equation}
$\gamma$ is approximately $1$ for no or weakly irradiated
atmospheres. However, in the case of strong irradiation,  $\gamma$
may differ significantly from unity. Moreover, it may be a strong
function of temperature, and, to a lesser extent, of density. This is easily seen by noting
that in optically-thin regions, the local mean intensity is essentially
equal to twice the irradiation intensity, since the incoming
intensity is equal to irradiation intensity, and the outgoing intensity is roughly
equal to it as well. The reason is that in order to conserve the
total flux when it is much smaller than the partial flux in the inward
or outward direction, both fluxes should be almost equal, and so too must 
the individual specific and mean intensities.

The local temperature in the upper layers is given, using 
equation (\ref{tupp}), as
\begin{equation}
\label{tupp2}
T/T_0 = \gamma(T)\, .
\end{equation}
It is clear that if $\gamma$ exhibits a strongly non-monotonic behavior
in the vicinity of $T_0$, for instance if it has a pronounced minimum or
maximum there, equation (\ref{tupp2}) may have two or even more
solutions.  

This is the origin of thermal inversions, when they occur, in the atmospheres 
of strongly irradiated EGPs (Hubeny, Burrows, \& Sudarsky 2003).  
The essential element is the differential absorption in the optical on 
the one hand (since this is where most of the irradiating stellar 
light is found), and in the infrared (since this is where most of the emission 
at the temperatures achieved in the atmospheres of hot Jupiters occurs).  
As the above formalism makes plain, an inversion is not possible for gray opacities, 
whatever the degree of irradiation.  For strongly irradiated EGPs, if there is not 
a strong optical absorber at altitude, then stellar optical light is absorbed rather 
deeply in the atmosphere, near pressures of $\sim$1 bar.  This is near and interior to 
the corresponding emission photospheres in the near-IR.  The result is 
a more-or-less monotonic decrease of temperature with altitude and decreasing 
pressure.  However, if there is a strong optical absorber at altitude, the large value 
of $\gamma$ will allow another solution to eq. (\ref{tupp2}) for which there is an inversion.
The resulting higher temperatures in the upper atmosphere will result in higher
fluxes in the mid-IR, e.g. in the IRAC and MIPS bands of {\it Spitzer}) where their
photospheres reside.  We see signatures of such inversions in the measured spectra
of many transiting EGPs, such as HD 209458b, XO-1, TrES-4, and HD 149026b.
However, to date we do not know what chemical species is absorbing at altitude in the 
optical (see \S{4.3}).

\bigskip
\textbf{ 3. LESSONS FROM JUPITER AND SATURN}
\bigskip

Jupiter and Saturn are the largest planets in our solar system and serve as initial
paradigms for the atmospheres of EGPs.  With the largest 
exoplanets invariably referred to as ``super-Jupiters,'' it is instructive to
start with an assessment of the properties of ``regular Jupiters.''  Both Jupiter and
Saturn have been studied with a variety of remote-sensing techniques across a wide spectral
range, and these approaches have provided sufficient information to determine physical and
chemical properties and their variation in both time and space.  In the case of Jupiter,
these physical and chemical properties have been bolstered by {\it in situ} observations
made by the Galileo mission atmospheric probe which both extended and served as a measure of
ground truth for remote-sensing observations.  This subsection will concentrate 
on properties of their atmospheres and the spectra of their upwelling fluxes 
in order to provide analogies with extrasolar planets 
which can boast fewer observational constraints.

Observations of Jupiter and Saturn have been made in some detail from the time of Galileo,
with serious ground-based observations taking place with the advent of photographic film and
spectroscopy.  In the second half of the 20th century, NASA spacecraft ventured to both -
first with the Pioneers 10 (Jupiter) and 11 (Jupiter and Saturn) flyby
spacecraft.  Later observations were made with more instrumentation (including an infrared
spectrometer, and wide- and narrow-angle cameras) by Voyagers 1 and 2, which visited both
planets, and provided an abundance of information on the dynamics, structure, and composition
of both Jupiter and Saturn.  The Galileo spacecraft orbited Jupiter and determined more about
the dynamics of individual regions on the planet and dropped a direct probe into its
atmosphere.  Most recently, the Cassini spacecraft obtained substantial information on the
atmosphere, flying by Jupiter in a gravity assist on its way to Saturn, which it began
orbiting in 2005.  Most recently the New Horizons spacecraft flew by Jupiter on its way to
Pluto, collecting information on Jupiter's clouds.

These spacecraft observations were complemented by observations from Earth-orbiting platforms
beginning with observations of auroral emission from Jupiter by the International Ultraviolet
Explorer (IUE).  They include ultraviolet through near-infrared imaging and spectroscopy by
the instrument complement on board the Hubble Space Telescope (HST), disk-averaged
spectroscopy by ESA's Infrared Space Observatory (ISO), and even X-ray observations by the
Chandra Observatory.  Additional spectroscopic and occultation observations were made
using instruments on board NASA's Kuiper Airborne Observatory (KAO).

The ever-improving competence of ground-based observational facilities has 
substantially increased our knowledge of these atmospheres. In particular, increasing
spatial and spectral discrimination became available with larger primary mirrors
(minimizing diffraction-limited spatial resolution) and a variety of active optical systems,
which have been extremely effective in reducing the blurring due to atmospheric turbulence.
These have been accompanied at the smaller-telescope end by a host of amateur Jupiter
watchers wielding increasingly sophisticated instrumentation, including multi-filtered CCD
cameras.

\bigskip
\noindent
\textbf{ 3.1 Spectra}
\bigskip

Models for the structure of the atmospheres of Jupiter and Saturn can be understood in terms
of the various regions in which energy transfer is by either radiation or convection, similar to
stellar atmospheres.  To good accuracy, much of the atmospheres of both planets can be
approximated as self-gravitating fluids in hydrostatic equilibrium. 
At depth, the atmospheres of both Jupiter and Saturn are dominated by convective processes, and gases rise and fall at
rates faster than they can radiate away energy.  Thus, adiabatic conditions
hold (eq. \ref{lapse}).  

It is important to note the ``principal players" in radiative
transfer in Jupiter and Saturn.  In the visible and near infrared, the spectrum is dominated
by CH$_4$ (methane) absorption, with some additional opacity from the H$_2$-H$_2$
collision-induced fundamental and H$_2$ quadrupole lines at higher resolution
Figure \ref{jupiter_spect} depicts the near-infrared spectrum of Jupiter from 1 to 6 microns. 
For Jupiter, the 5-$\mu$m spectrum is not dominated by reflected sunlight, as is the case at 
the shorter wavelengths, but by thermal emission from depth, as the result of a dearth of 
gaseous absorption in this region.  It is this atmospheric ``window'' which 
allows glimpses of Jupiter's composition down to a few bars of pressure, where H$_2$O 
vapor was first detected in Jupiter's atmosphere from observations using NASA's Kuiper
Airborne Observatory, and where high-resolution spectroscopy detected PH$_3$ and trace
constituents such as AsH$_3$ (arsine) and CO (carbon monoxide) at depth.  For colder Saturn,
more of this spectral window is composed of reflected sunlight.

At longer wavelengths, in the middle and far infrared, the spectrum is dominated by thermal
emission. Figure \ref{bright} shows the brightness temperature spectrum of both Jupiter and Saturn.
Rather than flux, these spectra are plotted in brightness temperature, which is defined
as the temperature that a blackbody source would need to emit at a particular wavelength
in order to match the observed flux.  

Plotting a spectrum in brightness temperature rather than directly in flux is a 
convenient way to display a spectrum over a broad range, such as the ones in Fig. \ref{bright}, because
the flux varies over several orders of magnitude.  In addition, plotting the brightness
temperature provides a quick way to estimate the depth in the atmosphere from which most of
the radiation emerges.  For example, the source of a strong emission feature with a high
brightness temperature is likely to be emerging from molecular emission very high in the hot
stratospheres of either planet.  More details can be found in dePater and Lissauer (2004), Irwin (2003),
or Goody and Yung (1989).

We also note that Fig. \ref{bright} plots its primary spectral scale in cm$^{-1}$ (``wavenumbers'')
rather than wavelength.  Historically, this is the result of such spectra being derived from
Fourier-transform spectrometers, where inverse of the path length difference of a
Michaelson-like interferometer is the native spectral unit.  A wavenumber (cm$^{-1}$) is
equal to the frequency of the radiation in Hertz (sec$^{-1}$) divided by 29.97.  Because the
frequency or wavenumber is proportional to quantum energy, such a plot enables molecular
lines to be displayed in a manner proportional to the transition energy.  The result is
often (but not always!) a regular series of line transitions (or groups of line transitions
known as manifolds).  Some of these are evident in Fig. \ref{bright} in the far infrared for NH$_3$
(ammonia) or PH$_3$ (phosphine) or near 10 $\mu$m for NH$_3$.  

Figure \ref{bright} shows that the far infrared, where the bulk of Jupiter's and Saturn's flux
emerges, is controlled by the collision-induced absorption of H$_2$ which varies so slowly
that it constitutes a virtual continuum opacity source.  Two broad rotational lines are
evident, and the far-infrared lines of NH$_3$ in Jupiter and PH$_3$ in Saturn show up as
absorption features on the long-wavelength ``translational'' component of the H$_2$
absorption.  When helium collides with H$_2$, a slightly different spectrum is created than
for pure H$_2$-H$_2$ absorption, and the translation component is stronger.  Fitting this
shape accurately provides a means to determine the ratio of He (helium) to H$_2$ in these
planets.  This part of the spectrum also allows one to determine the ratio of ortho-H$_2$, where the spins of
the hydrogen atoms are parallel, to para-H$_2$, where the spins are antiparallel (parallel
but pointed in opposite directions), because para-H$_2$ alone is responsible for all even
rotational transitions, such as S(0) and ortho-H$_2$ alone is responsible for all odd
rotational transitions, such as S(1).  In a
quiescent atmosphere, there is an equilibrium between these two species of H$_2$ which is
purely a function of temperature and how their energy levels are populated.  In reality,
however (particularly in Jupiter) H$_2$ gas can be transported vertically much faster than
this temperature re-equilibration can take place, and the para-ortho H$_2$ ratio varies
across the face of the planet. This is useful as an indirect means to track upwelling and
downwellling winds.  For example, a para-H$_2$ value which is lower than the value
expected from local thermal equilibrium, but greater than 0.25, would indicate rapid
upwelling from the deep, warm atmosphere where its value is close to the high-temperature
asymptotic value of 25\%.  However, at the high temperatures of hot-Jupiter atmospheres, transiting or otherwise,
para- and ortho-hydrogen are in thermal and statistical equilibrium.

The spectra of Jupiter and Saturn are also filled with discrete transitions, such as the rotational lines of
NH$_3$ and PH$_3$ in the far infrared.  Lines arising from a combination of vibrational
and rotational quantum transitions can be identified easily in the mid-infrared. Both planets
display emission features arising from CH$_4$ (methane) and from higher-order hydrocarbons,
such as C$_2$H$_2$ (acetylene) and C$_2$H$_6$ (ethane) which are by-products of the
photolysis of methane by ultraviolet radiation in the upper atmosphere.  They appear in
emission rather than absorption because the stratospheres of both planets are substantially
warmer than the upper troposphere.  This is the result of warming
by sunlight absorbed in the near infrared by CH$_4$ and small atmospheric particulates.
NH$_3$ and PH$_3$ can also be seen as components of the mid-infrared spectrum; their lines
are in absorption because they are most abundant in the troposphere.
In fact, ammonia is a condensate - much like water vapor in the Earth's atmosphere.  It does
not appear as strongly in Saturn's spectrum because Saturn is colder than Jupiter, and
ammonia condenses out deeper in the atmosphere than in Jupiter and is at the limits of
detectability above the H$_2$ collision-induced continuum.  PH$_3$ lines are, in fact, also
detectable in Jupiter's spectrum, but are more difficult to discern among the forest of
NH$_3$ lines in Fig. \ref{bright}.

The existence of absorption and emission features from H$_2$ and CH$_4$ in the thermal
spectrum is also useful for determining temperatures.  Both molecules are well mixed in the
atmosphere, so any changes which take place in their emission from point to point can
be attributed to changes of temperature rather than abundance.  There are
techniques for inverting the radiative-transfer equations for the emitted flux, given
observations of thermal emission in regions dominated by H$_2$ or CH$_4$ which allow
temperatures to be determined (some authors say ``retrieved'') as a function of altitude.
For the H$_2$ absorption features, the relevant range is in the upper troposphere - around
$\sim$100--400 mbar ($\sim$0.1--0.4 bars) in atmospheric pressure. For CH$_4$ emission, the relevant
range is the stratosphere at pressures less than $\sim$100 mbar.

Details about molecular transitions in planetary atmospheres can be found in Goody and Yung
(1989) and Irwin (2004).  Irwin (2004) also discuss in some detail the techniques for
retrieving atmospheric temperatures in the giant planets.

\bigskip
\noindent
\textbf{ 3.2 Clouds}
\bigskip

The visual appearances of Jupiter and Saturn are strongly affected by clouds.   
Jupiter's visible atmosphere is famously heterogeneous, and cloud motions and colors have been studied for
decades.  An extensive review of historical and amateur images is given by Rogers (1995).
Because light reflected from a generally homogeneous upper-atmospheric haze layer 
Voyager and Cassini imaging observations of Jupiter's and Saturn's cloud fields to determine wind velocities confirmed
and greatly refined ground-based studies which showed variable zonal (east-west) winds that 
are variable with latitude, but generally constant in time.  Jupiter's wind vectors are both
prograde and retrograde with respect to the rotation of the deep interior, ranging from a
maximum near 140 m/sec and a minimum near -50 m/sec.  Doppler tracking of the Galileo probe
suggests that these zonal winds persist into the atmosphere at least as deep at the 20-bar
pressure level. Saturn's winds display a strong jet of higher speeds, reaching 500 m/sec
near the equator.  A drop of Saturn's maximum jet speed has been detected in comparing
HST-derived winds with those derived by Voyager imaging, but it is not clear whether this is
due to a real change in the wind speed itself or a change of the altitude of the cloud
particles being tracked.  Meridional (north-south) winds are much smaller and only notable
around large, discrete features such as Jupiter's Great Red Spot or smaller vortices.
Jupiter's largest and longest-lived vortices are all anticylonic (rotating counterclockwise
in the southern hemisphere, clockwise in the northern hemisphere). Cyclonic storms are also
evident and Galileo observations noted that they are often the source locations for
observable lightning; however, their lifetimes are generally measured in weeks rather than
months to years.  

At near-infrared wavelengths, the dominance of gaseous absorption by CH$_4$ and some H$_2$
allows tracking of higher-altitude clouds and hazes.  Imaging through wavelengths that sample
particulates at various levels provides a means of shaping a 3-dimensional picture of cloud
systems.  Stratospheric hazes can be seen at high latitudes
which are thought to be products of charged-neutral interactions in Jupiter's polar auroral
regions.  Images of near-infrared reflectivity in Saturn have an overall banded structure
which is similar to Jupiter's, but Saturn seldom shows individual discrete features at these
wavelengths that are large enough to be detected from the Earth.

The 5-$\mu$m ``spectral windows'' for Jupiter and Saturn provide insight into their cloud
systems at several bars of atmospheric pressure.  The lack of substantial gaseous absorption
allows cloud tops to be sensed, and thermal emission begins to dominate most of the 4.8--5.3
$\mu$m region for Jupiter and the 5.0--5.3 $\mu$m region for Saturn.  Analogous to the
visible region, Jupiter's 5-$\mu$m appearance is spectacularly heterogeneous (Fig. \ref{jup5}), with
some regions of the atmosphere nicknamed ``5-$\mu$m hot spots'' ostensibly cloudless down to
deep (5 bars atmospheric pressure) and warm (275 K) cloud-top levels.  These are mixed with
regions of both intermediate and cold clouds across the face of the disk.  The warmer
regions are highly correlated with
areas of darker color in the visible, and the hot spots are correlated with gray areas.
This implies that the lighter colored clouds represent an upper layer of
tropospheric clouds with particles thick enough to be optically thick at wavelengths near 5
$\mu$m.  Such clouds could be related to an ammonia ice condensate (``cirrus'') cloud near
1 bar to 500 mbar in pressure, but spectral signatures of ammonia ice have only been seen in
particulates undergoing rapid upwelling or regions of
turbulent flow to the northwest of the Great Red Spot.  This requires that ammonia condensate
to be (i) riming nucleation sites made of other materials or (ii) particles covered by a
coating of other material.   The morphology of
these clouds can also be used to diagnose upwelling and downwelling motions, just as the
para-H$_2$ distribution discussed earlier.  Upwelling gas is wet and capable of forming
clouds, but - after reaching dryer and colder altitudes - the condensate has precipitated out
and downwelling regions will generally be cloudless because of the dry nature of the gas.
Thus the relatively cold 5-$\mu$m interior of the Great Red Spot (near 
the center of Fig. \ref{jup5}) diagnoses a region of upwelling, and its
ostensibly warm (actually less cloudy) periphery indicates regions of downwelling.

Saturn's thermal emission at $\sim$5 $\mu$m reveals a cloud structure in the several-bar
pressure region every bit as heterogeneous as Jupiter's, although one might say it is a
negative version.  A series of warm zonal bands are interlaced with a cold bands, and
discrete regions are dominated not by clear, warm areas as in Jupiter's atmosphere but by
cloudy, cold ones.  Cassini VIMS observations of this region are similar, but take
advantage of the entire spectral window by observing only thermal emisison on Saturn's night
side.  The persistence of some cloud features that are detectable in the visible can also be
found in this spectral region, an indicator that the dynamics creating them persists over a
wide vertical range.  For example an irregular band near high northern latitudes known as the
``polar hexagon'' attests to the presence of meridional waves which perturb the normally
uniform zonal flow, and the combination of visible, near-infrared, 5-$\mu$m, and mid-infrared
observations shows that this feature persists in both the cloud and thermal fields from
atmospheric pressures less than $\sim$100 mbar down to levels with pressures of several bars.

\bigskip
\noindent
\textbf{ 3.3 Temperatures}
\bigskip

Temperatures in the atmospheres of Jupiter and Saturn can be determined using several
different techniques.  For individual locations, occultations of optical light from stars by
the atmosphere has been used to determine atmospheric density as a function of altitude and
then transformed into temperatures and pressures using the ideal gas law, together with
the equation of hydrostatic equilibrium and an estimate of the mean molecular weight.  A similar approach has been
successful using occultations by the atmosphere of coherent spacecraft signals in the radio.

One such result for a Voyager spacecraft radio occultation is shown in Fig. \ref{jup_sat_profile} for Saturn, but
the Galielo probe at Jupiter provided the most direct measurement of temperatures available for the giant
planets (Fig. \ref{jup_sat_profile}), with direct measurements consistent with a ``dry'' adiabat (i.e., one
without perturbations to the lapse rate from exchanges of latent heat from condensing
molecules) from $\sim$400 mbar to $\sim$20 bars (thick, asterisked curve in Fig. \ref{jup_sat_profile}).  At higher levels,
before the entry velocity of the probe became sub-sonic, the deceleration of the probe was
Doppler tracked, the density of impeding gas determined, and knowledge of the mean molecular
weight was used to determine the temperature profile of Jupiter at that location in the
stratosphere (regular asterisked curve in Fig. \ref{jup_sat_profile}).  This profile showed the presence of
substantial vertical thermal waves in the stratosphere, which should provide an substantial
additional source of mechanical energy for the upper stratosphere.

The most successful approach for mapping temperatures across the disk is to invert
observations of thermal emission in spectral regions dominated by well-mixed constituents.  As
described earlier, spectral regions dominated by collision-induced absorption by well-mixed
H$_2$ and the 7-$\mu$m band of CH$_4$ (Fig. \ref{bright}) are used in both planets to 
map temperatures.  Fig. \ref{jup_sat_profile} contains temperatures derived using this technique by the Cassini Composite Infrared
Spectrometer (CIRS) instrument for Jupiter and Saturn near the equator and for latitudes of
$\pm$30$^\circ$.  This approach has a vertical resolution no better than a scale height, and
so it acts as a low-pass filter for temperature variability with altitude.  Nonetheless, it provides a good
means to derive a range of temperatures over time and space, similar to instrumentation in
weather satellites orbiting the Earth.  The technique has been expanded from spacecraft
spectroscopy to a series of middle-infrared filtered images of Jupiter and Saturn, 
which provide similar temperature information over a wide horizontal range without as
much vertical information overlap.  This has provided a means for ground-based observations
to supplement the close-up coverage provided by spacecraft, with spatial resolutions that
have often been as good as those available from spacecraft instruments with the
deployment of mid-infrared imaging instruments on 8-meter or larger-class telescopes which
minimize the effects of blurring by diffraction.  
The entire host of temperature information verifies that the atmosphere is indeed separable
into largely convective versus largely radiative regions, with the convective regions over a
variety of different latitudes and bands converging at depth in both planets (Fig. \ref{jup_sat_profile}).

\bigskip
\noindent
\textbf{ 3.4 Compositions}
\bigskip

Over 99.9\% of the composition of both planets by volume consists of hydrogen, helium and
methane.  A summary of our current knowledge of the compositions of Jupiter and Saturn in the
detectable part of the atmosphere is given in Table 1.  Their relative abundances,
particularly the He/H$_2$ ratio, are important constraints on models of their formation,
evolution and interiors.  For Jupiter, the value shown in Table 1 is the one derived
independently by the Galileo Probe Mass Spectrometer (GPMS) and a purpose-built Helium
Abundance Detector (HAD).  The measured value for the He abundance may be influenced by
its ability to form droplets at high pressure and ``rain out'' of the deep atmosphere toward
the center of the planet. Neon (Ne) is soluble in the He drops and appears to be severely
depleted in Jupiter with respect to solar abundances.  The abundances of Ar (argon), Kr
(krypton) and Xe (xenon) determined by the GPMS are all enhanced by $\sim$2.6 with respect
to solar values.  

CH$_4$ is present throughout the atmosphere because it represents the simplest stable form
of carbon, originally delivered to the planet as methane or other carbon-bearing ices in
planetestimals.  This is also true of NH$_3$ for N, H$_2$O for O, and possibly PH$_3$ for P,
and H$_2$S for S.  Ammonia and water both condense out in the detectable parts of Jupiter's
and Saturn's atmospheres; both are responsible for at least part of the atmospheric opacity
in the submillimeter through microwave part of the spectrum and,  with phosphine, in the 5-$\mu$
spectral window.  The measurement of O/H via the Galileo GPMS value is considered to be too
low because the Galileo probe descended in a 5-$\mu$m ``hot spot,'' that also proved to be a region of unusual
dessication.  Chemical equilibrium models predict that H$_2$S should react with NH$_3$ to
form NH$_4$SH (ammonium hydrosulfate) clouds around $\sim$1$-$3 bars pressure (between the upper
NH$_3$ cloud and the deeper H$_2$O cloud layers.  Molecules in chemical disequilibrium at the
cold upper tropospheric temperatures such as PH$_3$ and GeH$_4$ (germane) are present because
upwelling from warmer depths takes place more rapidly than molecular decomposition.

Methane is present in the stratosphere and is destroyed by solar
ultraviolet radiation, with higher-order hydrocarbons resulting from the chemical
recombination process.  The most abundant of these are C$_2$H$_2$ (acetylene), C$_2$H$_4$
(ethylene), and C$_2$H$_6$ (ethane), which act as the principal means for stratospheric
radiative cooling.  Enhanced abundances of
hydrocarbons have been reported in polar stratospheric ``hot spot'' regions, the result of
additional auroral-related chemistry.  Reports of H$_2$O line emission attest to the presence
of H$_2$O in the stratosphere, in abundances suggesting infall from the exterior, either from
ring particles or from interplanetary ice.

The isotopic $^{13}$C/$^{12}$C ratio, determined by the Galileo probe for Jupiter and by the
Cassini CIRS experiment for Saturn are close to the terrestrial value, suggesting a
protosolar value with almost no chemical fractionation of carbon isotopes in either
atmosphere.  The observed D/H ratios, determined from the same set of experiments, also
suggests protosolar values.  In the case of Saturn, this is inconsistent with interior models
which predict an enhancement of deuterium because of the larger mass of its core.

In addition to the reviews already mentioned, detailed summaries on several topics 
concerning Jovian and Saturnian atmospheres, interiors, icy satellites, rings and 
magnetospheres are given in Bagenal et al. (2004) for Jupiter and Brown et al. (2009) for Saturn.

\bigskip
\centerline{\textbf{ 4. SUMMARY OF EGP FEATURES AND HIGHLIGHTS}}
\bigskip

Using the formalisms sketched in \S{2}, and the current knowledge of Jovian planet 
atmospheres summarized in \S{3}, one can establish a theoretical framework 
in which to study EGPs for any age, orbit, mass, and elemental composition.
This theoretical edifice is also understood using the simple models provided in \S{2.4} 
and \S{2.6}, with which one can understand most of their measured and anticipated properties.
In this section, we highlight a few of the interesting ideas and features that have emerged
in the last few years concerning EGP evolution, generic planet/star flux ratios, and
the distinctive character of the family of hot Jupiters. This summary is in no way 
comprehensive, but provides the reader with a snapshot of some of the salient facts and issues 
in this rapidly evolving field.

\bigskip
\noindent
\textbf{ 4.1 Evolution of EGP Atmospheres and Chemistry}
\bigskip

Unlike in a star, EGP atmospheric temperatures are sufficiently low that
chemistry is of overriding importance.  The atmosphere of a gaseous
giant planet is the thin outer layer of molecules that controls its
absorption and emission spectra and its cooling rate.  Molecular hydrogen (H$_2$) is the
dominant constituent, followed by helium.  An EGP's effective temperature (\teff) can vary from $\sim$2500 K at birth for
the more massive EGPs or in a steady state for the most severely irradiated to $\sim$50 K for the least
massive EGPs in wide orbits after evolving for billions of years.   This wide range translates into a wide variety
of atmospheric constituents that for a given mass and elemental composition can evolve
significantly.

Without any significant internal sources of energy, after formation an EGP
gradually cools and shrinks.  Its rate of cooling can be altered by stellar irradiation,
or when old and light by hydrogen/helium phase separation (Fortney \& Hubbard 2004).
Jupiter itself is still cooling and its total infrared
plus optical luminosity is about twice the power absorbed from the Sun.
Hence, the temperatures and luminosities achieved are not just functions of
mass and composition, but of mass, age, composition, orbital distance, and stellar type.

Since EGPs evolve through atmospheres of various compositions
and temperatures, age is a key parameter in their study.  A theoretical
evolutionary scenario for Jupiter itself can serve to exemplify these transformations.
The evolution of other EGPs with different masses and orbits will be different
in detail, but for the wide-separation variety not in kind. The description
below is based in part on the review by Burrows (2005), to which the reader is referred.
Another useful review can be found in Marley et al. (2007).

At birth, Jupiter had a \teff\ near 600-1000 K and the appearance
of a T dwarf (Burgasser et al. 2002).  It had no ammonia or water clouds and, due to the presence
of atomic sodium in its hot atmosphere, had a magenta color in the optical (Burrows et al. 2001).
Due to the formation and settling to depth of the refractory silicates that condense in
the temperature range $\sim$1700-2500 K, its atmosphere
was depleted of calcium, aluminum, silicon, iron, and magnesium
\footnote{Note that such silicate clouds should exist at depth in the current Jupiter.}.
Water vapor (steam) was the major molecule containing oxygen, gaseous methane was the major reservoir of carbon,
gaseous ammonia and molecular nitrogen were the contexts for
nitrogen, and sulfur was found in H$_2$S. At lower temperatures ($< \sim 700$K), FeS
would be the equilibrium reservoir of sulfur. However, as noted above, earlier in
the planet's history when refractory species condensed out and settled
gravitationally, the atmosphere was left depleted of most
metals, including iron.  The result of this ``rainout" was that
the chemistry simplified and sulfur was in the form of H$_2$S,
as is observed in the current Jupiter. The rainout phenomenon of condensates
in the gravitational field of EGPs and solar system giants is a universal feature
of their atmospheres (Fegley \& Lodders 1994; Burrows \& Sharp 1999).

As Jupiter cooled, the layer of alkali metals was buried below the photosphere to higher pressures, but
gaseous H$_2$, H$_2$O, NH$_3$, and CH$_4$ persisted. At a \teff\ of $\sim$400 K, water
condensed in the upper atmosphere and water clouds appeared.
This occurred within its first 100 million years.  Within less than a billion years, when
\teff\ reached $\sim$160 K, ammonia clouds emerged on top of the water clouds,
and this layering persists to this day (see section 3).

Stellar irradiation retards cloud formation, as does a large EGP mass,
which keeps the EGP hotter longer. Proximity to a star also keeps the
planet hotter longer, introducing a significant dependence
of its chemistry upon orbital distance. Around a G2V star such as
the Sun, at 5 Gyr and for an EGP mass of 1.0 \mj, water clouds form
at 1.5 AU, and ammonia clouds form beyond 4.5 AU (Burrows,
Sudarsky, \& Hubeny 2004).  Jupiter's and Saturn's current effective
temperatures are 124.4 K and 95 K, respectively. Jupiter's orbital
distance and age are 5.2 AU and 4.6 Gyr. The orbital distance, mass, and
radius of a coeval Saturn are 9.5 AU, 0.3 \mj, and 0.85 \rj.  However, as an EGP
of whatever mass cools, its atmospheric composition evolves through a similar
chemical and condensation sequence.  Figure \ref{profiles} depicts the atmospheric
temperature/pressure (T/P) profile for a sequence of 1-\mj, 5-Gyr models
as a function of orbital distance (0.2 to 15 AU) from a G2V star.  As
the EGP's orbital distance increases, its atmospheric temperature at a
given age and pressure decreases. Superposed on the plot are the H$_2$O
and NH$_3$ condensation lines.  Clearly, a given atmospheric composition
and temperature can result from many combinations of orbital distance,
planet mass, stellar type, and age.  This lends added complexity to
the study of EGPs.

The atmospheres of hot Jupiters at orbital distances of $\sim$0.02-0.07
AU from a G, F, or K star are heated and maintained at temperatures of $\sim$1000-2000 K,
roughly independent of planet mass and composition.  The transiting EGPs
discovered to date are examples of such hot objects.
At high temperatures, carbon is generally in the form of carbon monoxide, not methane.
As a result of this and the rainout of most metals, EGP atmospheric compositions are
predominantly H$_2$, He, H$_2$O, Na, K, and CO. There are, however, significant
day/night differences in composition, temperature, and spectrum that distinguish
a hot Jupiter from a lone and isolated planet or star.  For instance, on the night side,
carbon might be found in methane, whereas on the day side it might be in CO.  This
suggests that non-equilibrium chemistry, where the chemical rates and the dynamical motions
compete in determining the composition, might be at play.
Exotic general circulation models (GCMs) with credible
dynamics, radiative transfer, chemistry, and frictional effects will soon be necessary
to understand the equatorial currents, jet streams, day/night differences, terminator
chemistry, and global wind dynamics of irradiated EGPs, in particular,
and of orbiting, rotating EGPs, in general.

There is another interesting aspect to the orbital-distance dependence of
EGP properties and behavior.  At the distance of Jupiter, the infrared photosphere
is close to the radiative-convective boundary, near 0.5 bars.  This means
that a good fraction of the stellar radiation impinging upon Jupiter from the Sun
is absorbed directly in the convective zone, which thereby redistributes this
heat more or less uniformly to all latitudes and longitudes of the planet.
The upshot of this, and that the luminosity from the core due to the remaining
residual heat of formation is comparable to the stellar irradiation,
is that Jupiter (and Saturn) emit isotropically in the infrared,
the effects of cloud banding and structure not withstanding.  However, as
a gas giant planet moves inward towards its primary star, the pressure level
of the photosphere and the pressure level of the day-side radiative-convective
zone separate.  When the EGP is at $\sim$0.05 AU, and after a Gyr, the
photospheres have moved little, but the radiative-convective boundary is now
near a kilobar.  Moreover, the internal flux is miniscule compared with the magnitude
of the irradiation.  The result is that heat is not efficiently redistributed
by internal convective motions, but by zonal winds in the radiative atmosphere
{\em around} the planet to the night side.  This day side/night side dichotomy
is a central feature of hot Jupiters and can lead to severe thermal contrasts
as a function of longitude.

Curiously, a star with solar compostion at the edge of the light-hydrogen-burning
main sequence (\mstar$\sim$75 \mj) has a \teff\ of $\sim$1700 K.
Therefore, an irradiated EGP, with a radius comparable to that of such a star,
can be as luminous.

\bigskip
\noindent
\textbf{ 4.2 Planet/Star Flux ratios of Wide-Separation EGPs}
\bigskip

The planet/star flux ratio versus wavelength is the key quantity
is the study of orbiting planets. Figure \ref{contrastd}, taken
from Burrows, Sudarsky, \& Hubeny (2004),
depicts orbit-averaged (Sudarsky et al. 2000) such flux ratios from
0.5 \mic to 30 \mic for a 1-\mj/5-Gyr EGP in a circular orbit at distances of 0.2 to 15 AU
from a G2V star.  These models are the same as those depicted in Fig. \ref{profiles}.
The water absorption troughs are in evidence throughout.  For the closer EGPs at
higher atmospheric temperatures, carbon resides in CO and methane features are weak.
For those hot Jupiters, the Na-D line at 0.589 \mic and the corresponding resonance
line of K I at 0.77 \mic are important absorbers, suppressing flux in the visible
bands.  Otherwise, the optical flux is increased by Rayleigh scattering
of stellar light.   As $a$ increases, methane forms and the methane
absorption features appear in the optical (most of the waviness
seen in Fig. \ref{contrastd} for $a$ \sgreat\ 0.5 AU shortward of 1 \mic), at
$\sim$3.3 \mic, and at $\sim$7.8 \mic. At the same time, Na and K disappear from the atmosphere
and the fluxes from $\sim$1.5 \mic to $\sim$4 \mic drop.  For all models, the
mid-infrared fluxes longward of $\sim$4 \mic are due to self-emission, not reflection.
As Fig. \ref{contrastd} makes clear, for larger orbital distances a separation between a
reflection component in the optical and an emission
component in the mid-infrared appears.  This separation into components
is not so straightforward for the closer, more massive, or younger family members.
For these EGPs, either the large residual heat coming from the core
or the severe irradiation buoys the fluxes from 1 to 4 \mic.  The more massive
EGPs, or, for a given mass, the younger EGPs, have larger $J$, $H$, and $K$ band fluxes.
As a result, these bands are diagnostic of mass and age.  For EGPs with large orbital distances,
the wavelength range from 1.5 \mic to 4 \mic between the reflection and emission components
may be the least favorable search space, unless the planet is massive or young.

When water or ammonia clouds form, scattering off them enhances the optical
fluxes, while absorption by them suppresses fluxes at longer wavelengths in, for example, the 4$-$5 \mic window.
Because water and ammonia clouds form in the middle of this distance sequence,
the geometric albedo ($A_g$) is not a monotonic function of $a$.
This effect is demonstrated in Figure \ref{fig_ratiodist}, taken from Sudarsky et al. (2005),
which plots the planet/star flux ratio for a Jovian-mass planet orbiting a G2V central star
as a function of orbital distance at 0.55 $\mu$m, 0.75 $\mu$m, 1 $\mu$m, and 1.25 $\mu$m.
Clearly, the planet/star flux ratio does not follow an inverse-square law.
Clouds can form, evaporate, or be buried as the degree of stellar heating varies with
varying distance. This behavior is included in the models of Fig. \ref{contrastd}, but its precise
manifestations depend upon unknown cloud particle size, composition, and patchiness.
As a consequence, direct spectral measurements might constrain cloud properties.
As Fig. \ref{contrastd} suggests, the planet/star contrast ratio is better
in the mid- to far-infrared, particularly at wide separations.  For such separations, the
contrast ratio in the optical can sink to 10$^{-10}$.

For the closest-in EGPs (not shown on Fig. \ref{contrastd}),
such as HD 189733b, HD209458b, OGLE-TR56b, 51 Peg b, and $\tau$ Boo b, the contrast
ratio in the optical should be between 10$^{-5}$ and 10$^{-6}$ and is more favorable,
though still challenging.  However, for these hottest EGPs, alkali metal (sodium
and potassium) absorption is expected to dominate in the optical and very near-IR.
Sodium has already been detected in the transit spectrum of HD 209458b. As a result, the
expectation is that the optical albedos of hot Jupiters should be quite low,
perhaps less than $\sim$5\%.

Importantly, the transiting EGPs are so hot and their orbits are so favorably inclined
that the planet/star flux ratios in the mid-IR, in particular in the {\em Spitzer} bands,
can range between $\sim$10$^{-3}$ and $\sim$10$^{-2}$.  For these objects, the gap between the
reflection and emission components is completely closed.  Moreover, the variation in the summed light
of planet and star during eclipses and the execution of the orbit can be used to derive
the planetary spectrum itself.  

\bigskip
\noindent
\textbf{ 4.3 Hot Jupiter Highlights}
\bigskip

The discovery of $\sim$60 (and counting) transiting giant planets
allows one to address their physical structures by
providing simultaneous radius and mass meansurements.  However, 
the proximity of transiting EGPs to their primaries boosts
the planet/star flux contrast ratios in the near- and mid-IR 
to values accessible by {\em Spitzer} during secondary eclipse. 
Such measurements in the IRAC ($\sim$3.6, $\sim$4.5, $\sim$5.8, 
$\sim$8 $\mu$m) and MIPS ($\sim$24 $\mu$m) bands, as well as 
IRS spectral measurements from $\sim$5 to $\sim$ 14 $\mu$m, are 
the first {\it direct} detections of planets outside the solar system.  
In addition, {\em Spitzer} has yielded photometric light curves as a 
function of orbital phase, and revealed or constrained brightness distributions 
across planet surfaces.  This has been done (dramatically so) for HD 189733b 
at 8 and 24 microns, for HD 149026b and GJ 436b at 8 microns, and, though 
non-transiting, for both $\upsilon$ And b at 24 microns and HD 179949b 
at 8 microns. It will soon be done for many more hot Jupiters. 
Excitingly, {\em Spitzer} measurements alone have yielded exoplanet 
compositions, temperatures, and longitudinal temperature variations
that have galvanized the astronomical and planetary communities. 

Furthermore, using NICMOS, STIS, and ACS on HST, precision measurements of transit 
depths as a function of wavelength have enabled astronomers to identify atmospheric 
compositions at planet terminators.  In this way, sodium, water, methane, hydrogen, 
and carbon monoxide have been inferred.  Moreover, the MOST micro-satellite has
obtained a stringent upper limit of $\sim$8\% to the optical albedo of HD 209458b (Rowe 
et al. 2008).  Hence, in ways unanticipated just a few years ago, these space telescopes 
are constraining the degree of heat redistribution from the day to the night sides 
by zonal winds (providing a glimpse of global climate), are signaling the presence of 
thermal inversions, and are revealing exoplanet chemistry. The new 
data from {\em Spitzer}, HST, and MOST have collectively inaugurated the era of 
remote sensing of exoplanets and the techniques articulated in this chapter were 
developed to interpret such data.

As described above, there are many recent highlights in the study of giant exoplanet
atmospheres, but we will focus here on only a few examples.  They are the spectral and photometric
measurements of HD 189733b by Grillmair et al. (2008) and Charbonneau et al. (2008) 
and the photometric measurements of HD 209458b by Knutson et al. (2008) and Deming et al. 
(2005), along with their interpretations (see also Barman 2008). These objects 
exemplify the two basic classes of EGP atmospheres which have emerged, those 
without and with significant thermal inversions at altitude, and are the two best 
studied transiting EGPs.

The top panel of Fig. \ref{hd189.209.spectra} shows a comparison of the HD 189733b data
with three theoretical models for the planet/star flux ratio. The best fit model (black) is consistent 
with most of the data and assumes solar elemental abundances and only modest heat redistribution to the 
night side (P$_n$ = 0.15).  One of the salient features is the fact that the IRAC 1 ($\sim$3.6 $\mu$m)
to IRAC 2 ($\sim$4.5 $\mu$m) ratio is greater than one.  This has been interpreted to
mean that the temperature profile monotonically decreases outward and that there is no significant
inversion. The left panel of Fig. \ref{tp.eps} shows such temperature profiles (but for TrES-1), 
as well as the locations of the effective photospheres in the {\em Spitzer} bands.  An appreciable inversion would 
reverse the IRAC 1/IRAC 2 ratio, since the effective photosphere in IRAC 2 is further out and at much 
lower pressures than that for IRAC 1.  In addition, the depression of the IRAC 2 flux is 
consistent with the presence of carbon monoxide, which has a strong absorption 
feature there.  More important is the comparison between theory and the IRS spectrum
of Grillmair et al. (2008) between $\sim$5 and $\sim$14 microns.  Not only is the 
slope of the spectrum roughly matched, but there is a $\sim$3-$\sigma$ detection of a feature just longward of $\sim$6 microns.
This is best interpreted as the flux peak due to the opacity window between the P and R branches of the $\nu_2$ vibrational
bending mode of water vapor (see Fig. \ref{fig1.sharp}) .  Water would explain both the general slope of the IRS spectrum 
and this peak and together they are now taken to be the best indications of the presence of water (steam) in
an EGP atmosphere. 

The bottom panel of Fig. \ref{hd189.209.spectra} compares theoretical
models for the planet/star flux ratio of HD 209458b at secondary eclipse 
with the corresponding {\em Spitzer} photometric data of Knutson et. 
al. (2008) (IRAC) and Deming et al. (2005) (MIPS).  The differences 
with HD 189733b are illuminating.  Here, the IRAC 1/IRAC 2 ratio is less than 
one and the IRAC 3 ($\sim$5.8 $\mu$m) flux is higher than the IRAC 4 ($\sim$8 $\mu$m)
flux.  These features are best interpreted as signatures of
a thermal inversion at altitude.  There must be an absorber of optical (and/or near UV) stellar
light to create the thermal inversion and a hot outer atmosphere 
(see right panel of Fig. \ref{tp.eps}), with the result that the 
planet/star ratio is much higher in the IRAC 1, 2, and 3 channels 
than predictions without inversions (e.g., the black curve without an absorber).  
In fact, centered near IRAC 3, the IRAC data look like an {\em emission} 
feature, centered in the broad water band from $\sim$4.5 to $\sim$8 $\mu$m, 
as opposed to the absorption trough expected for an atmosphere with a 
negative temperature gradient. Emission features are expected in 
atmospheres with thermal inversions.   

HD 189733b and HD 209458b together represent the two classes of hot-Jupiter atmospheres
that are emerging to be explained.  As Fig. \ref{hd189.209.spectra} indicates, 
we can reproduce their basic spectral features, but there remain many anomalies.  The fits
at $\sim$24 $\mu$m are problematic and the detailed shapes of the spectra are not well reproduced.
Moreover, the models depend upon the degree of heat redistribution to the night side (modeled 
as P$_n$; Burrows, Budaj, \& Hubeny 2008), which crucially depends upon the unknown
zonal flows and global climate.  More important, though we know that there are inversions in the atmospheres of 
HD 209458b, TrES-4, XO-1b, and (likely) HD 149026b, we don't know why.  What is the nature
of this ``extra absorber" in the optical (\S{2.6})?  With what other planetary properties
is the presence or absence of an inversion correlated?  Metallicity?  Gravity? The substellar flux on the planets
TrES-1 and XO-1b are similar, but only the latter shows an inversion. In the bottom panel of 
Fig. \ref{hd189.209.spectra}, we have introduced an ad hoc optical absorber at altitude
merely to determine the opacity ($\kappa_e$) needed to reproduce the measurements. Its
physical origin is not addressed.  Some have suggested it is TiO or VO (Fortney et al. 2008).  
However, these species can easily condense out and should not persist 
for long in the upper atmosphere.  The so-called cold-trap, which operates
in the Earth's upper atmosphere to render it dry, should also operate for these metal oxides
here (Spiegel, Silverio, \& Burrows 2009).  Hence, the nature of the extra optical absorber at altitude is currently unknown.
One might well ask whether the slight thermal inversion introduced due to zonal
heat redistribution at depth can be the culprit (Hansen 2008), but the magnitude of the resulting
temperature excursion is slight ($\le 100$ K) and does not approach the $\sim$1000 K
needed to explain the data. 
A recent suggestion by Zahnle et al. (2009) involves
the photolytic production of allotropes of sulfur and HS.
%
%
However, whether a sulfur allotrope or any sulfur compound achieves the
necessary abundance at the requisite upper pressure levels and what its production
systematics are with the properties of the planet and star (metallicity and/or stellar
spectrum) remains to be seen. The theoretical task is emerging to be complex and challenging.

\bigskip
\centerline{\textbf{ 5. FUTURE PROSPECTS}}
\bigskip

The theoretical study of exoplanet atmospheres in general, and of EGP atmospheres in particular,
has been energized by continuing discoveries and surprises which show no sign of abating.  
In the short term, the last cold cycle and the upcoming warm cycle(s) of {\em Spitzer}
will provide more light curves and secondary eclipse measurements.  HST will continue
to yield high-sensitivity transit data.  Ground-based extreme adaptive optics systems
that will enable high-contrast imaging of exoplanets from under the glare of their parent
stars will be developed and perfected.  In the next decade, JWST will constitute a 
quantum leap in sensitivity and broader wavelength coverage, which together may lay to 
rest many of the outstanding questions which have emerged during the early heroic phase of 
EGP atmospheric characterization. These include, but are not limited to:   

\begin{itemize}

\item How do zonal flows in EGP atmospheres affect their spectra and phase curves?

\item What is the magnitude of heat redistribution to the planet's night side? What is ``$P_n$"? 

\item How bright is the night side?

\item Upon what do the longitudinal positions of the hot and cold spots of
strongly irradiated planets depend?

\item What is the depth of heat redistribution?  

\item What causes the strong thermal inversions in a subset of the hot Jupiters?

\item Is there a central role for photolysis and non-equilibrium chemistry in EGP 
atmospheres, particularly in hot Jupiter atmospheres?  

\item Can the chemistry and irradiation be out of phase? Is there chemical and thermal hysteresis?

\item How different are the compositions on a planet's day side and night side?

\item How different are the temperatures and chemical compositions 
at the terminators of hot Jupiters during ingress and egress?

\item Could there be significant gaps in our understanding of the dominant chemical 
constituents of EGP atmospheres?

\item What is the origin of dissipation in planetary atmospheres that
stabilizes their zonal flows?

\item What are the signatures of non-zero obliquity and asynchronous spin 
in the measured light curves and spectra?  

\item What are the unique features of the light curves of EGPs in highly elliptical orbits and 
can tidal heating contribute to their thermal emissions?

\item What are the thermal and chemical relaxation time constants of EGP 
atmospheres under the variety of conditions in which they are found?

\item At what orbital distances are the day-side and night-side IR fluxes 
the same due to efficient heat redistribution? 

\item What is the role of metallicity and planet mass in EGP compositions, light curves, and spectra?

\item What are the spectral and evolutionary effects of atmospheric clouds and hazes? 

\item Do the convective cores of hot Jupiters cool differently on their night and day sides?

\end{itemize}

Our theories are poised to improve significantly,
with credible multi-dimensional simulations undertaken and non-equilibrium chemistry 
and cloud formation addressed.  Methods are being designed to understand day-night coupling,
the degree of heat redistribution, and the pattern of zonal flow. Moreover, 
with the discovery of more transiting planets such as GJ 436b in the ice 
giant mass and radius regime, the variety of signatures will multiply.
Finally, more wide-separation EGPs will be separately imaged and probed.
Given the brilliant nature of its past, whatever the manifold uncertainties, this field
is on a trajectory of future growth and excitement that is establishing it as a core discipline
of 21st-century astronomy.

\bigskip
\textbf{ Acknowledgments.} The authors would like to acknowledge Ivan Hubeny, David 
Sudarsky, and Christopher Sharp for past collaborations and much 
wisdom.  They would also like to thank Travis Barman and Dave Spiegel for careful 
readings of the manuscript and numerous constructive suggestions.  This work 
was partially supported by NASA via Astrophysics Theory 
Program grant \# NNX08AU16G and with funds provided under 
JPL/Spitzer Agreements No. 1328092 and 1348668. GSO would like to
acknowledge partial support by funds to the Jet Propulsion
Laboratory, California Institute of Technology.  We are grateful to NASA's Infrared
Telescope Facility and JPL collaborators Leigh Flethcher and Padma Yanamandra-Fisher for the
image shown in Fig. \ref{jup5}.

\bigskip

\centerline\textbf{ REFERENCES}
\bigskip
\parskip=0pt
{\small
\baselineskip=11pt

\refs Bagenal, F., Dowling, T. E., and McKinnon, W. B. (eds.) (2004) {\em Jupiter: The
Planet, Satellites and Magnetoshere} Cambridge University Press, Cambridge, UK.

\refs Barman, T. (2008) On the Presence of Water and Global Circulation 
in the Transiting Planet HD 189733b. {\em Ap.~J.}, 676, L61-64.

\refs Brown, R., Dougherty, M., Krimigis, S. (eds.) (2009) {\em Saturn After Cassini/Huygens}.
Springer, Heidelberg.

\refs Burgasser, A. et al. (2002). The Spectra of
T Dwarfs. I. Near-Infrared Data and Spectral Classification.
{\em Ap.~J.}, 564, 421-451.

\refs Burrows, A. \&
Sharp, C.M. (1999) Chemical Equilibrium Abundances in Brown Dwarf and Extrasolar Giant Planet Atmospheres.
{\em Ap.~J.}, 512, 843-863.

\refs Burrows, A., Hubbard, W.B., Lunine, J.I.,
\& Liebert, J. (2001) The theory of brown dwarfs and extrasolar giant planets.
{\em Rev. Mod. Phys.}, 73, 719-765.

\refs Burrows, A., Sudarsky, D., \& Hubeny, I. (2004) Spectra and Diagnostics
for the Direct Detection of Wide-Separation Extrasolar Giant Planets.
{\em Ap.~J.}, 609, 407-416.

\refs Burrows, A. (2005) A theoretical look at the direct detection of giant planets outside the Solar System.
{\em Nature}, 433, 261-268.

\refs Burrows, A., Budaj, J., \& Hubeny, I. (2008) Theoretical Spectra and Light Curves of Close-in
Extrasolar Giant Planets and Comparison with Data. {\em Ap.~J.}, 678, 1436-1457.

\refs Burrows, A., Ibgui, L., \& Hubeny, I. (2008) Optical Albedo Theory of 
Strongly-Irradiated Giant Planets: The Case of HD 209458b. {\em Ap.~J.}, 682, 1277-1282.

\refs Charbonneau, D., Knutson, H.A., Barman, T., Allen, L.E., Mayor, M., Megeath, 
S.T., Queloz, D., \& Udry, S.  (2008) The broadband infrared emission spectrum of the
exoplanet HD 189733b. {\em Ap.~J.}, 686, 1341-1348.  


\refs Deming, D., Seager, S., Richardson, L.J., \& Harrington, J. (2005)
Infrared radiation from an extrasolar planet.
{\em Nature}, 434, 740-743.

\refs dePater, I. and Lissauer, J.J. (2004). {\em Planetary Sciences} Cambridge University
Press, Cambridge, UK.

\refs Dyudina, U. A., Sackett, P. D., Bayliss, D. D, Seager, S.,
Porco, C. C., Throop, H. B., \& Dones, L. (2004)
Phase light curves for extrasolar Jupiters and Saturns.
\verb"astro-ph/0406390".

\refs Fegley, B. \& Lodders, K. (1994) Chemical models of the deep atmospheres of Jupiter and Saturn.
{\em Icarus}, 110, 117-154.

\refs Fegley, B. \& Lodders, K. (2001) Very High Temperature Chemical Equilibrium Calculations with the CONDOR Code.
{\em Meteoritics \& Planetary Science}, 33, A55.

\refs Fortney, J.J. \& Hubbard, W.B. (2004).
Effects of Helium Phase Separation on the Evolution of Extrasolar Giant Planets.
{\em Ap.~J.}, 608, 1039-1049.

\refs Fortney, J.J. (2005) The effect of condensates on the characterization 
of transiting planet atmospheres with transmission spectroscopy. {\em M.N.R.A.S.}, 364, 649-653.

\refs Fortney, J.J., Lodders, K., Marley, M.S., \& Freedman, R.S. (2008)
A Unified Theory for the Atmospheres of the Hot and Very Hot Jupiters: Two Classes of Irradiated Atmospheres.
{\em Ap.~J.}, 678, 1419-1435.

\refs Freedman, R.S., Marley, M.S., \& Lodders, K. (2008) 
Line and Mean Opacities for Ultracool Dwarfs and Extrasolar Planets.
{\em Ap.~J.~Suppl.}, 678, 1419-1435.

\refs Goody, R. M. and Yung, Y. L. (1989). {\em Atmospheric Radiation: Theoretical Basis, 2nd
Edition}. Oxford University Press, New York.

\refs Grillmair, C., Burrows, A., Charbonneau, D., Armus, L.,
Stauffer, J., Meadows, V., van Cleve, J., von Braun, K., \&
Levine, D. (2008) Strong Water Absorption in the Dayside Emission Spectrum of the Exoplanet
HD 189733b. {\em Nature}, 456, 767-769.

\refs Hansen, B.M.S. (2008) On the Absorption and Redistribution of Energy in Irradiated Planets. 
{\em Ap.~J.\ Suppl.}, 179, 484-508.

\refs Hubeny, I. (1990) Vertical structure of accretion disks - A simplified analytical model.
{\em Ap.~J.}, 351, 632.

\refs Hubeny, I., Burrows, A., \& Sudarsky, D. (2003) Possible Bifurcation
in Atmospheres of Strongly Irradiated Stars and Planets.
{\em Ap.~J.}, 594, 1011-1018.

\refs Irwin, P. G. J., Weir, A. L., Taylor, F. W., Lambert, A. L., Calcutt, S. B.,
Cameron-Smith, P. J., Baines, K., Orton, G. S., Encrenaz, T., and Roos-Serote. M. (1998)
Cloud structure and atmospheric composition of Jupiter retrieved from Galileo near-infrared
mapping spectrometer real-time spectra. {\em J. Geophys. Res. 103}, 23001-23021.

\refs Irwin, P. G. J. (2003) {\em Giant Planets of Our Solar System: Atmospheres, Composition
and Structure}, Praxis, Chichester, UK.

\refs Karkoschka, E. (1999) Methane, Ammonia, and Temperature
Measurements of the Jovian Planets and Titan from CCD-Spectrophotometry.
{\em Icarus}, 133, 134-146.

\refs Knutson, H.A., Charbonneau, D., Allen, L.E., Burrows, A., \& Megeath, S.T. (2008)
The 3.6-8.0 $\mu$m Broadband Emission Spectrum of HD 209458b: Evidence for an Atmospheric Temperature Inversion.
{\em Ap.~J.}, 673, 526-531.

\refs Chevallier, L., Pelkowski, J. \& Rutily, B. (2007) Exact Results in 
Modeling Planetary Atmospheres $-$ I. Gray Atmospheres.
{\em J.Q.S.R.T.}, 104, 357-376.

\refs Lodders, K. (1999) Alkali Element Chemistry in Cool Dwarf Atmospheres.
{\em Ap.~J.}, 519, 793-801.

\refs Lodders, K. \& Fegley, B. (2002)
Atmospheric Chemistry in Giant Planets, Brown Dwarfs, and
Low-Mass Dwarf Stars. I. Carbon, Nitrogen, and Oxygen.
{\em Icarus}, 155, 393-424.

\refs Marley, M.S., Gelino, C., Stephens, D., Lunine, J.I., \& Freedman, R. 
(1999) Reflected Spectra and Albedos of Extrasolar Giant Planets. I. Clear and Cloudy Atmospheres.
{\em Ap.~J.}, 513, 879-893.
 
\refs Marley, M.S., Fortney, J., Seager, S., \& Barman, T. (2007) 
Atmospheres of Extrasolar Giant Planets. In {\em Protostars and Planets V}, 
(B. Reipurth, D. Jewitt, and K. Keil eds.), University of Arizona 
Press, Tucson, pp. 733-747.


\refs Perryman, M.A.C. (2003) GAIA Spectroscopy: Science and Technology.
in ASP Conference Proceedings, Vol. 298
(ed. Ulisse Munari), p. 3 (ASP Conf. Series, 2003).

\refs Rogers, J. H. (1995) {\em The Giant Planet Jupiter}, Cambridge Univ. Press, Cambridge,
UK.

\refs Rowe, J.F. et al. (2008) The Very Low Albedo of an Extrasolar Planet: MOST Spacebased Photometry of HD 209458.
{\em Ap.~.J.}, 689, 1345-1353.

\refs Seager, S., Whitney, B.A, \& Sasselov, D.D. (2000)
Photometric Light Curves and Polarization of Close-in Extrasolar Giant Planets.
{\em Ap.~J.} 540, 504-520.

\refs Sharp, C.M. \& Huebner, W.F. (1990) Molecular equilibrium with condensation. {\em Ap.~J.~Suppl.}, 72, 417-431.

\refs Sharp, C.M. \& Burrows, A. (2007) Atomic and Molecular Opacities
for Brown Dwarf and Giant Planet Atmospheres. {\em Ap.~J.~Suppl.}, 168, 140-166.

\refs Sobolev, V. V. 1975, {\it Light Scattering in Planetary Atmospheres},
(Oxford: Pergamon Press Ltd.)

\refs Spiegel, D., Silverio, K., \& Burrows, A. (2009) 
Can TiO Explain Thermal Inversions in the Upper Atmospheres of
Irradiated Giant Planets?. {\em to Ap.~J.}, 699, 1487-1500  

\refs Sudarsky, D., Burrows, A.,
\& Pinto, P. (2000) Albedo and Reflection Spectra of Extrasolar Giant Planets.
{\em Ap.~J.}, 538, 885-903.

\refs Sudarsky, D., Burrows, A., \& Hubeny, I. (2003). Theoretical Spectra and 
Atmospheres of Extrasolar Giant Planets. {\em Ap.~J.}, 588, 1121.

\refs Sudarsky, D., Burrows, A., Hubeny, I., \& Li, A. (2005). Phase
Functions and Light Curves of Wide-Separation Extrasolar Giant Planets.
{\em Ap.~J.}, 627, 520-533.

\refs Unwin, S.C. \& Shao, M. (2000) Space Interferometry Mission.
in {\it Interferometry in Optical Astronomy} (eds. P. J. Lena
\& A. Quirrenbach), 754-761.

\refs van de Hulst, H.C. (1974) The Spherical Albedo of a Planet Covered with 
a Homogeneous Cloud Layer. {\em Astron. \& Astrophys.}, 35, 209-214.

\refs Zahnle, K., Marley, M.S., Lodders, K., \& Fortney, J.J. (2009) 
Atmospheric Sulfur Photochemistry on Hot Jupiters. {\em Ap.~J.}, 701, L20-24  

\newpage

\begin{deluxetable}{cccccc}
\tabletypesize{\small}
\tablecaption{Compositions of Jupiter and Saturn in Volume Mixing Ratios\label{table1}}
\tablewidth{0pt}
\tablehead{Gas & Jupiter & Saturn\\ }
\startdata
H$_2$ & 0.862$\pm$0.003 & 0.882$\pm$0.024 \\
He & 0.136$\pm$0.003 & 0.113$\pm$0.024 \\
CH$_4$ & 2.4$\pm$0.5 x 10$^{-3}$ & 4.7$\pm$0.2 x 10$^{-3}$ \\
NH$_3$ & $\le$ 7 x 10$^{-4}$ & $\le$ 1 x 10$^{-4}$  \\
H$_2$O & $\le$ 6.5$\pm$2.9 x 10$^{-5}$ & 0.2--2 x 10$^{-7}$ \\
CO$_2$ &                               & $\le$ 3 x 10$^{-10}$ \\
CO & 1 x 10$^{-9}$  & (1--3) x 10$^{-9}$ \\
PH$_3$ & $\le$ 6 x 10$^{-6}$ & $\le$ 4.5 x 10$^{-6}$ \\
GeH$_4$ & $\sim$2 x 10$^{-8}$ & 3.4 x 10$^{-7}$ \\
H$_2$S & 1--8 x 10$^{-5}$ & $<$10$^{-6}$ \\
C$_2$H$_2$ & $\le$ 3 x 10$^{-8}$ & $\le$ 1 x 10$^{-7}$ \\
C$_2$H$_6$ & $\le$ 3 x 10$^{-6}$ & $\le$ 5 x 10$^{-6}$ \\
HCN & $<$ 1 x 10$^{-10}$ & $<$ 10$^{-9}$ \\
CH$_3$D & $\sim$2 x 10$^{-8}$ & 3.2$\pm$0.2 x 10$^{-7}$ \\
HD & $\sim$1.1 x 10$^{-5}$ & 0.7-1.7 x 10$^{-8}$ \\
Ar & 1.0 $\pm$ 0.4 x 10$^{-5}$ &  & \\
Ne & 2.3 $\pm$ 0.25 x 10$^{-5}$ &  & \\
\enddata
\end{deluxetable}

\newpage

\begin{figure*}
 \epsscale{2.0}
 \plotone{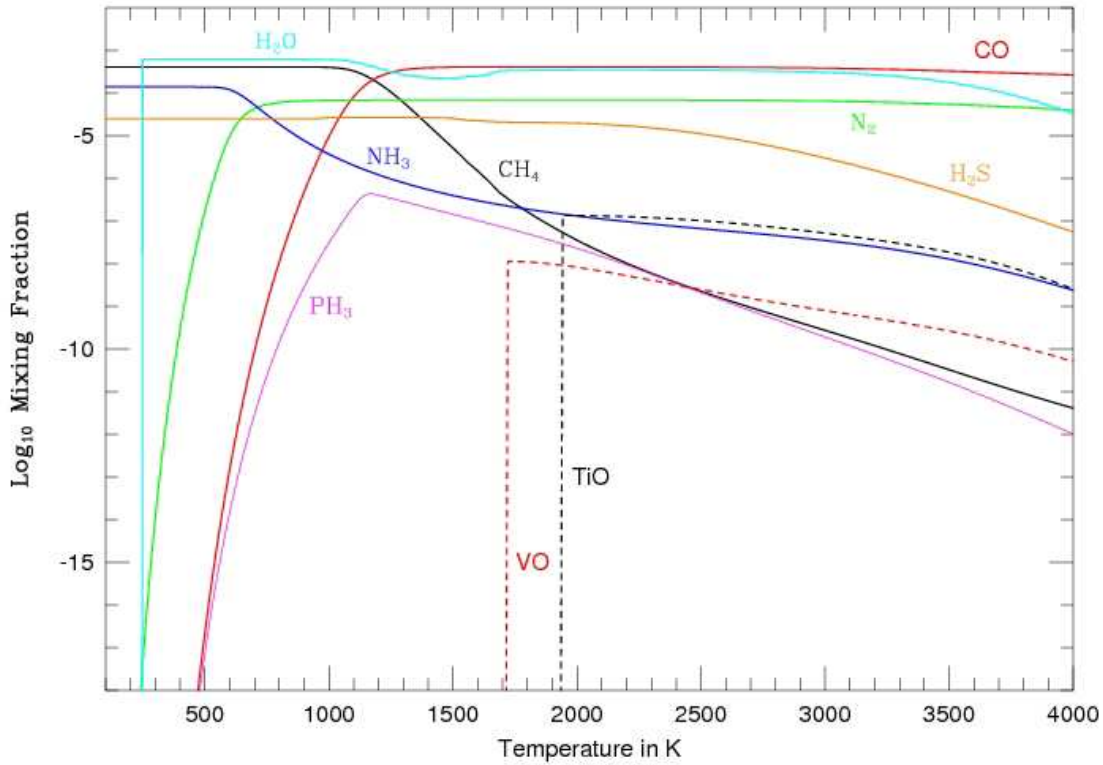}
 \caption{\small The log (base 10) of the mixing fraction as a function of
temperature at a total gas pressure of 1 atmosphere for the seven molecules
shown with solid curves $CH_4$ (black), $CO$ (red), $N_2$, $NH_3$, $H_2O$, $H_2S$, and
$PH_3$, and the two molecules shown with dashed curves
$TiO$ and $VO$.  At 4000 K,
$CO$ and $N_2$ are the most stable species, containing nearly all
the carbon and nitrogen, respectively.  With decreasing temperature,
$CO$ reacts with $H_2$, forming $CH_4$, which becomes the dominant
carbon-bearing species at low temperatures, and $N_2$ reacts with
$H_2$, forming $NH_3$, which likewise becomes the dominant
nitrogen-bearing species at low temperatures.  Except above about 3000 K,
$H_2O$ is fully associated containing nearly all the available oxygen
that is not bound in $CO$.  Below about 1600 K, its abundance
temporarily falls slightly due to the condensation of silicates which
reduce the available oxygen; however, the mixing fraction of $H_2O$
then rises again when $CO$ is converted to $CH_4$, which releases
the oxygen tied up in $CO$.  Finally, at 273 K $H_2O$ drops
effectively to zero due to the condensation of ice.
With decreasing temperature, both
$TiO$ and $VO$ rise as they associate, then sharply drop to
effectively zero when condensates involving $Ti$ and $V$ form.
[Taken from Sharp \& Burrows (2007)]
\label{f17.sharp}}
 \end{figure*}

\newpage

\begin{figure*}
 \epsscale{2.0}
 \plotone{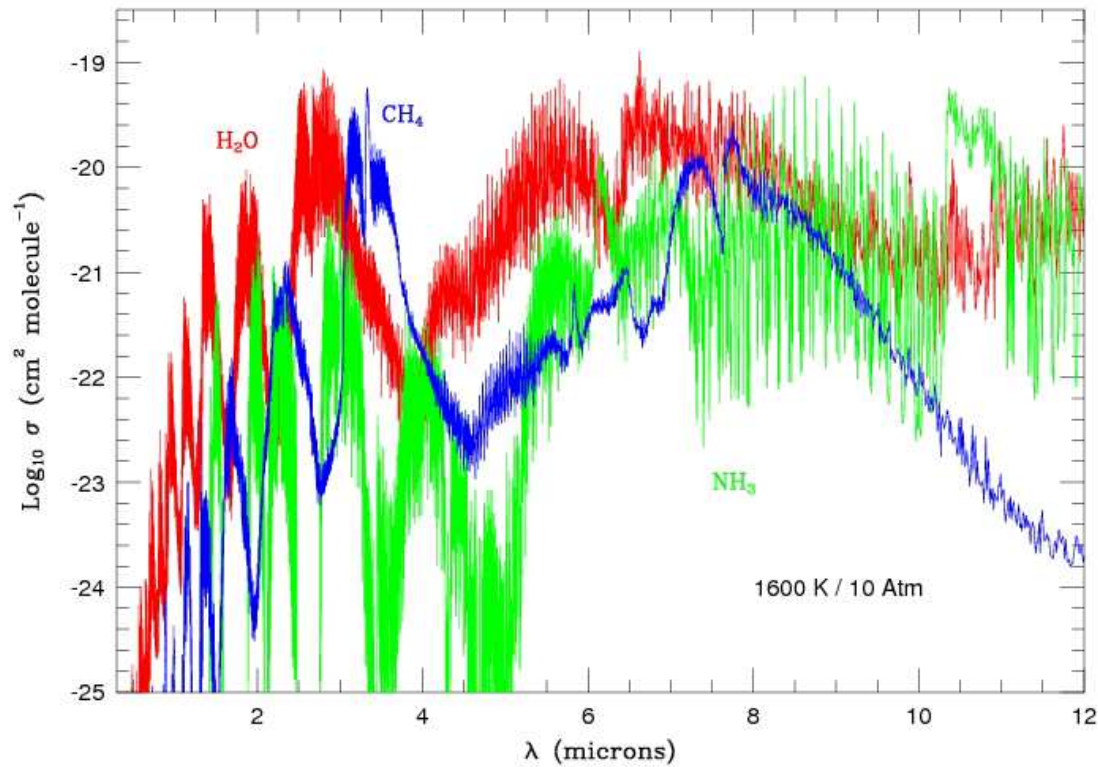}
 \caption{\small The log (base 10) of the monochromatic absorption $\sigma$ in
cm$^2$molecule$^{-1}$ as a function wavelength $\lambda$ in
$\mu$m in the infrared at a temperature of 1600 K and a pressure of 10
atmospheres for the vibration-rotation transitions of $H_2O$, $NH_3$, and $CH_4$.
The contribution due to different isotopes is included.
For this plot we chose a high enough temperature and pressure to ensure that the lines
were sufficiently broadened to suppress very rapid and large-amplitude
fluctuations in the absorption cross section that can otherwise be in evidence over short wavelength intervals.
In this way, the main band features (which are nevertheless generic
for each species) are be more easily seen.  At significantly
lower pressures the broadening of the lines is much smaller and the
absorption can change so rapidly in short wavelength intervals that the
main features do not show up so clearly.  As can be seen here, $H_2O$ has
a strong absorption feature just shortward of 3 $\mu$m, and $CH_4$ has a
strong peak near of 3.3 $\mu$m.  In the region of 8 $\mu$m to 9 $\mu$m
all three molecules absorb strongly; however, between about 10.5 $\mu$m
and 11 $\mu$m $NH_3$ has absorption which is distinctly higher than that
of the other two molecules.  When the combined opacity is calculated, the
individual absorptions must be weighted by the abundances.
[Taken from Sharp \& Burrows (2007)]
\label{fig1.sharp}}
 \end{figure*}


\begin{figure*}
 \epsscale{1.5}
 \plotone{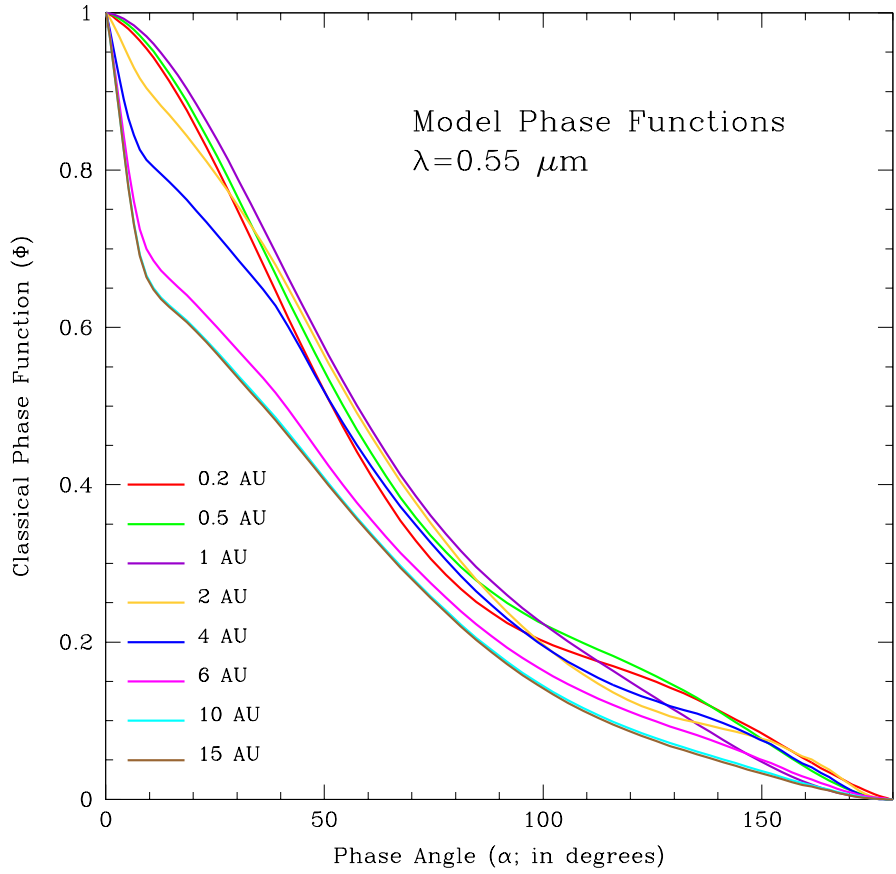}
 \caption{\small Theoretical optical phase functions of 1-\mj, 5-Gyr
EGPs ranging in orbital distance from 0.2 AU to 15 AU from a G2V star.  Near full phase, the
phase functions for our baseline models at larger orbital distances peak most strongly.
For the cloud-free EGPs at smaller
orbital distances (0.2 AU, 0.5 AU, and 1 AU), the phase functions are more rounded near
full phase. [Taken from Sudarsky et al. (2005)]
\label{fig_phaseall}}
 \end{figure*}

\newpage

\begin{figure*}
 \epsscale{1.5}
 \plotone{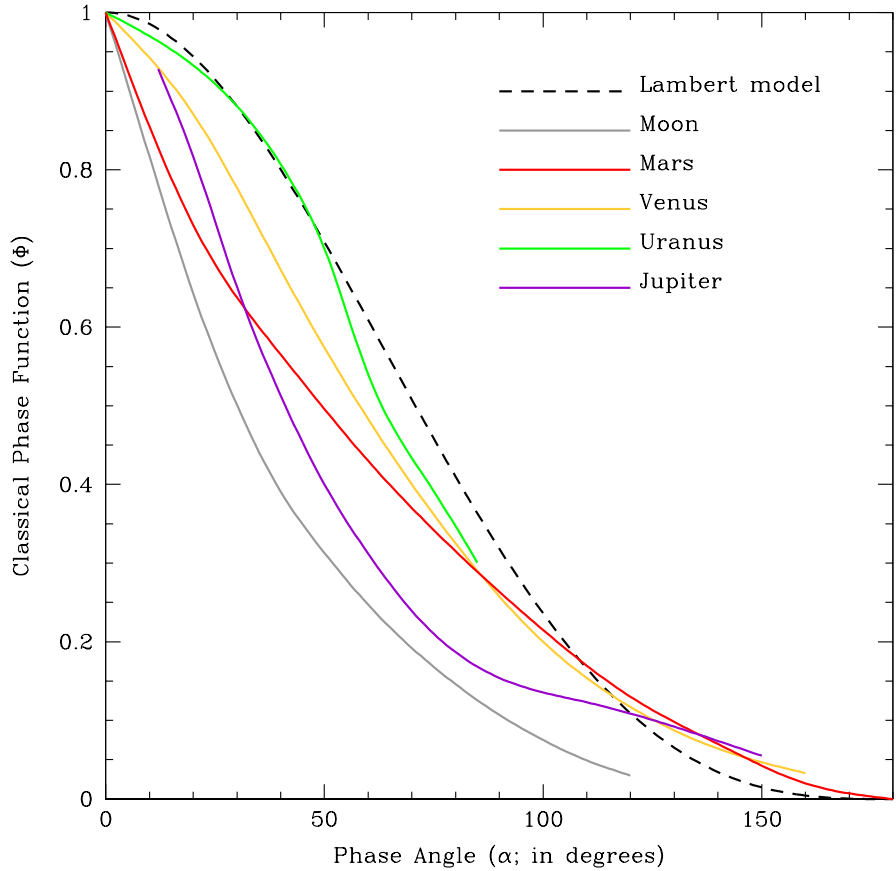}
 \caption{\small The measured visual phase functions for a selection of Solar System objects.
A Lambert scattering phase curve,
for which radiation is scattered isotropically off the surface regardless of
its angle of incidence, is shown for comparison.  The phase functions of the Moon
and Mars peak near full phase (the so-called ``opposition effect'').  A red bandpass Jupiter
phase function, taken from Dyudina et al. (2004), is also plotted.
[Taken from Sudarsky et al. (2005)]}
\label{realphase}
 \end{figure*}

\newpage


\newpage

\begin{figure*}
\includegraphics[width=12cm,angle=270]{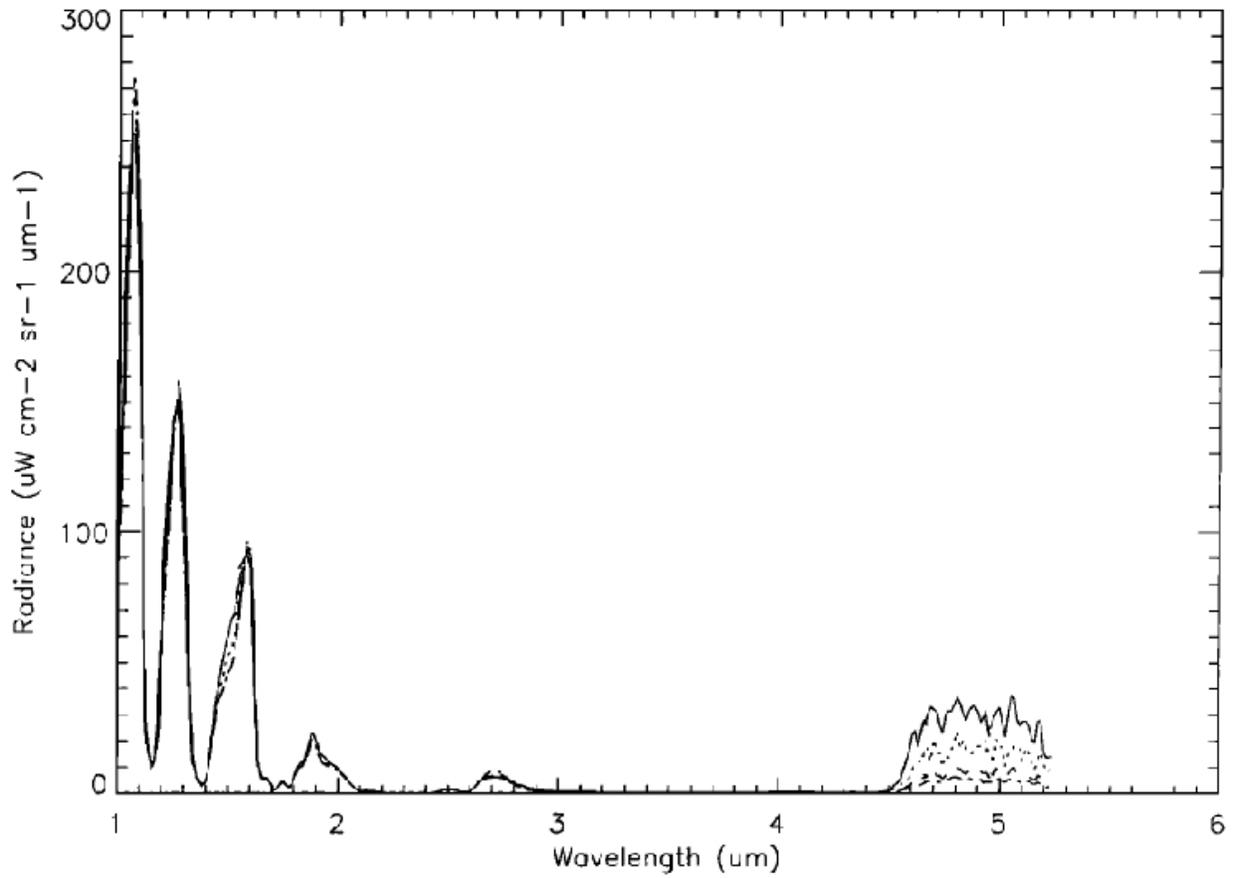}
 \caption{\small Near-infrared spectrum of Jupiter taken in a relatively ``cold'' but
reflective region of the atmosphere by the Galileo Near-Infrared Mapping Spectrometer (NIMS)
in 1996.  All radiances except those near 5 $\mu$m are sunlight reflected from clouds.
Absorption is primarily due to CH$_4$ gas with some contributions from a H$_2$
collision-induced fundamental band near 2.2 $\mu$m. From Irwin {\it et al.} (1998).
\label{jupiter_spect}}
\end{figure*}

\begin{figure*}
\includegraphics[width=12cm,angle=90]{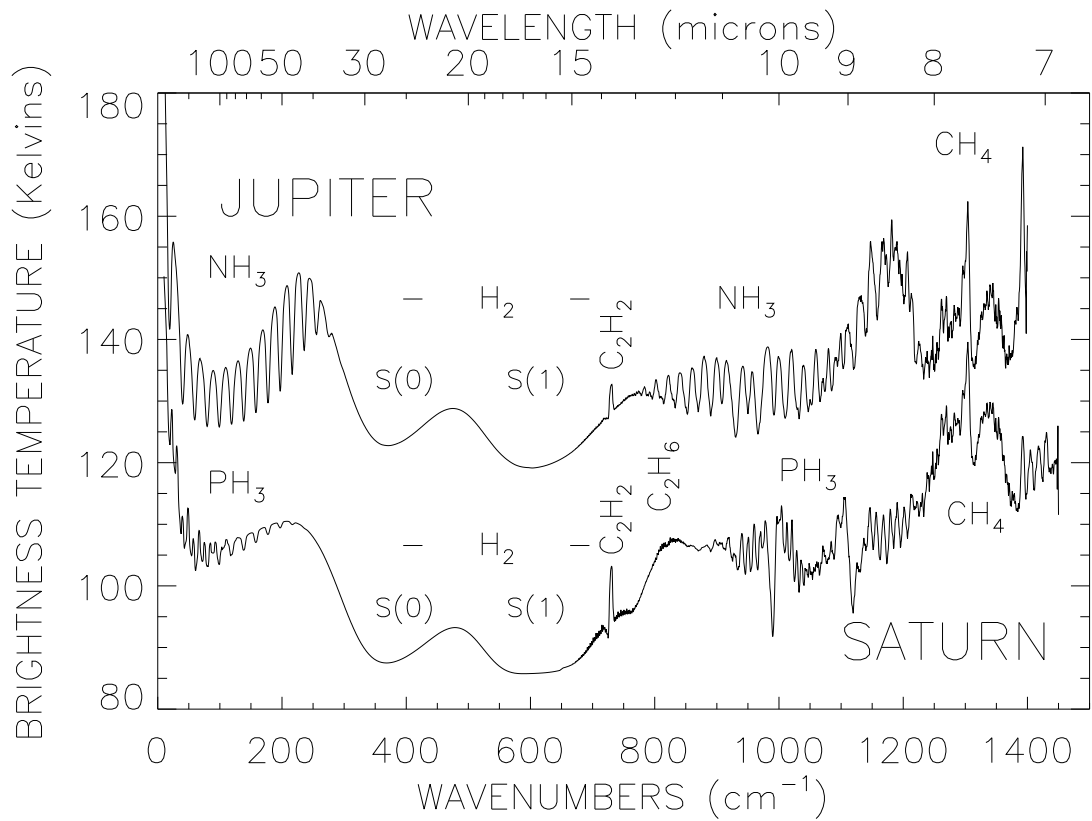}
 \caption{\small Middle-infrared brightness temperature spectra of Jupiter and Saturn,
derived from Voyager and Cassini infrared spectrometers.  Prominent spectral features
are identified.  Note that the H$_2$ collision-induced absorption is a smooth continuum
in the spectrum, with the broad rotational transitions S(0) and S(1) identified.
\label{bright}}
\end{figure*}

\newpage

\begin{figure*}
\centerline{\includegraphics[width=13cm,angle=270]{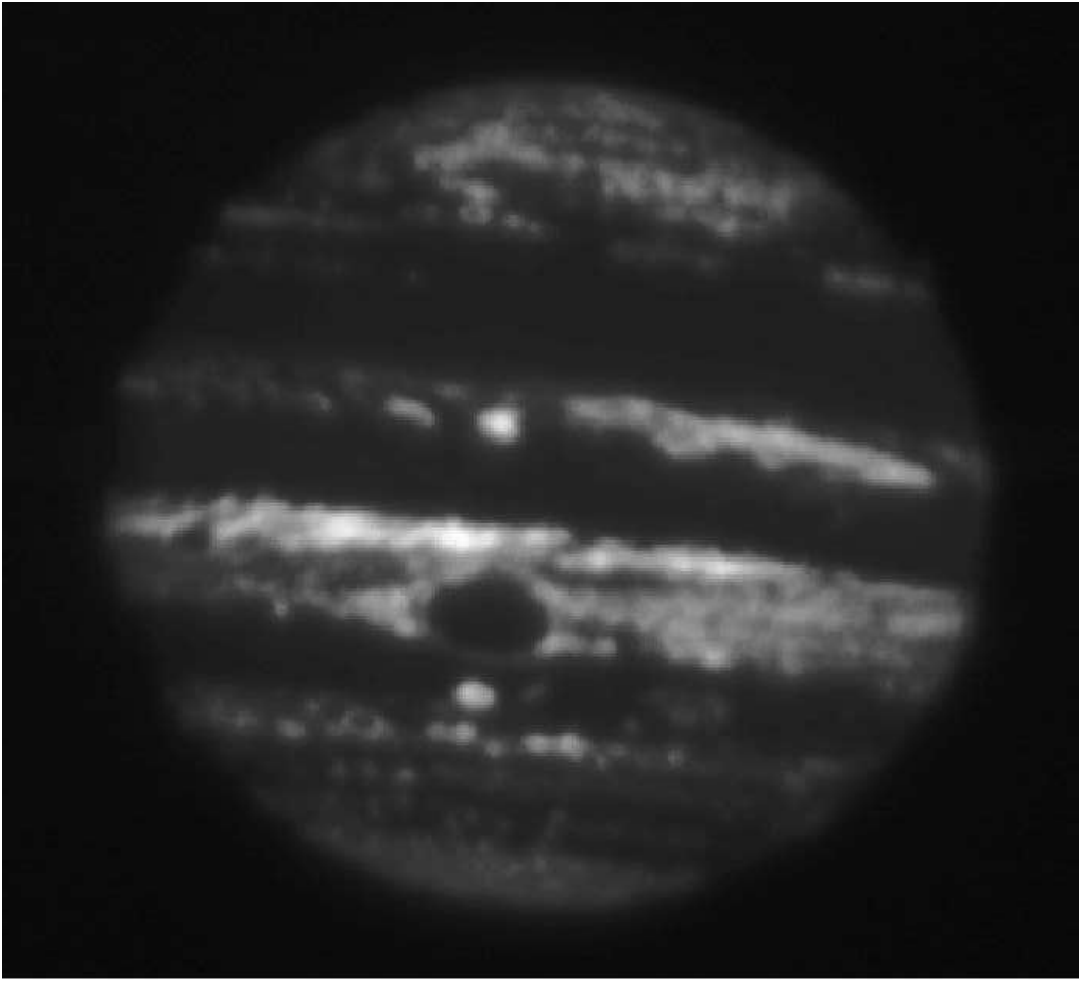}}
 \caption{\small 4.85-$\mu$m image of Jupiter taken at the NASA Infrared
Telescope Facility using the facility NSFCam2 instrument on 2008 Sept 24.
The filter used is 0.24 $\mu$m wide and illustrates the primarily thermal
radiance emerging from Jupiter's multi-layered cloud systems, with more
radiance emerging from cloud tops in the deepest atmosphere.  Jupiter's
Great Red Spot can be seen in the lower center of the figure as a region
whose periphery is defined by dry, downwelling gas which creates a region
relatively clear of clouds.
\label{jup5}}
\end{figure*}

\begin{figure*}
\includegraphics[width=12cm,angle=90]{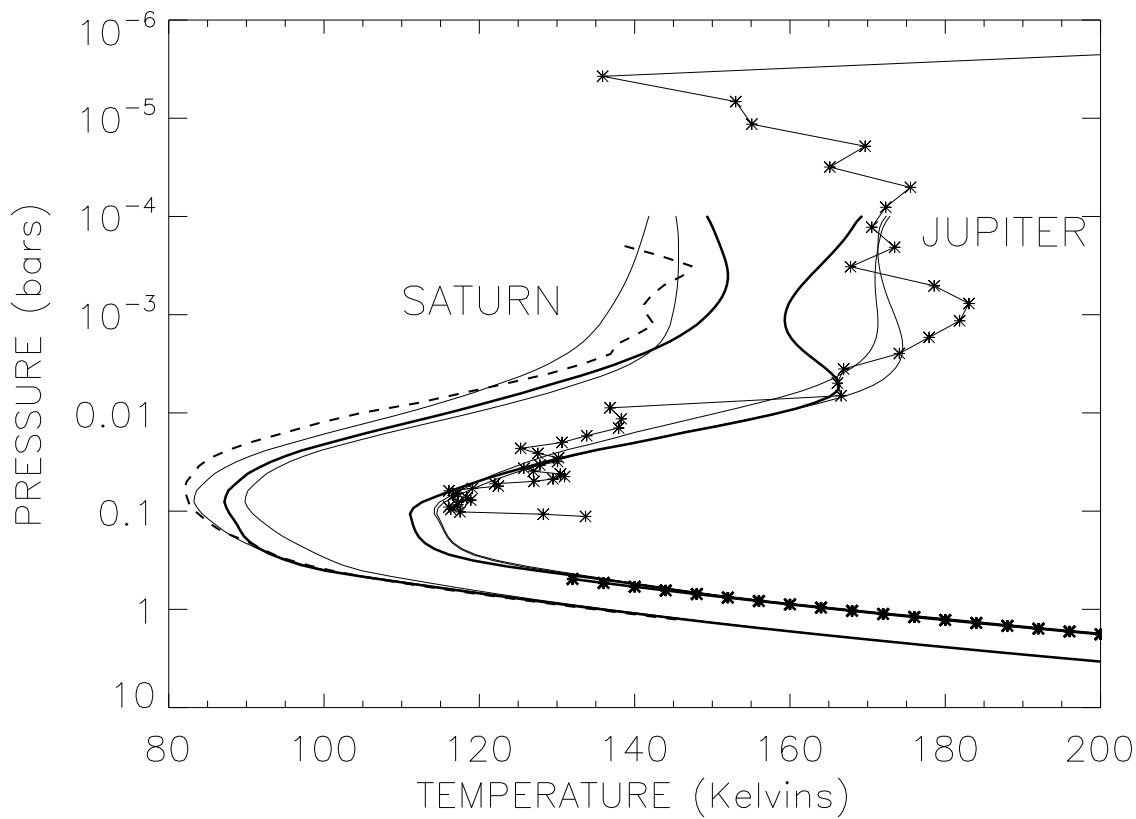}
 \caption{\small  Temperature profiles for Jupiter and Saturn. Curves in solid lines are
derived from Cassini CIRS (infrared) experiment observations, representative of
near-equatorial (thick solid) and $\pm$30$^\circ$ latitude (regular solid) curves.  Galileo
probe Atmospheric Structure Instrument (ASI) results are shown by the regular solid curve and
asterisks for the accelerometer data in the ``inverted'' stratosphere and by the thick solid
curve and asterisks for the direct measurements in the troposphere.  A Voyager radio
occultation curve is shown for Saturn by the dashed curve.  Note the warmer temperatures in
Jupiter than in Saturn, the ``inverted'' temperatures in the radiatively controlled
atmosphere at pressures lower than $\sim$100 mbars, vertical waves measured by the Galileo
ASI accelerometer experiment, and the convergence of the several temperature profiles at
different latitudes with depth.
\label{jup_sat_profile}}
\end{figure*}

\begin{figure*}
 \epsscale{1.5}
 \plotone{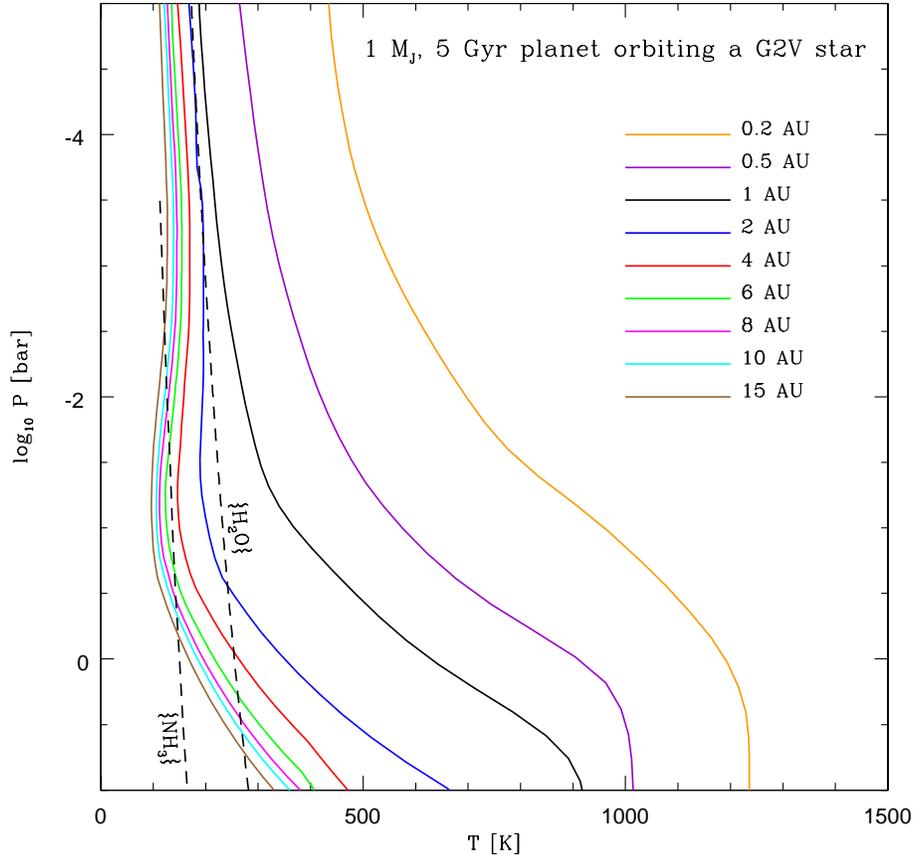}
 \caption{\small Profiles of atmospheric temperature (in Kelvin) versus the logarithm base ten of the pressure (in bars)
for a family of irradiated 1-\mj EGPs around a G2V star as a function of orbital distance.
Note that the pressure is decreasing along the ordinate, which thereby resembles altitude.
The orbits are assumed to be circular, the planets are assumed to
have a radius of 1 \rj, the effective temperature of the inner boundary flux is
set equal to 100 K,  and the orbital separations vary from 0.2 AU to 15 AU.
The intercepts with the dashed lines identified with either \{NH$_3$\} or \{H$_2$O\} denote the
positions where the corresponding clouds form.
[Taken from Burrows et al. (2004)]
\label{profiles}}
 \end{figure*}

\newpage

\begin{figure*}
 \epsscale{1.5}
 \plotone{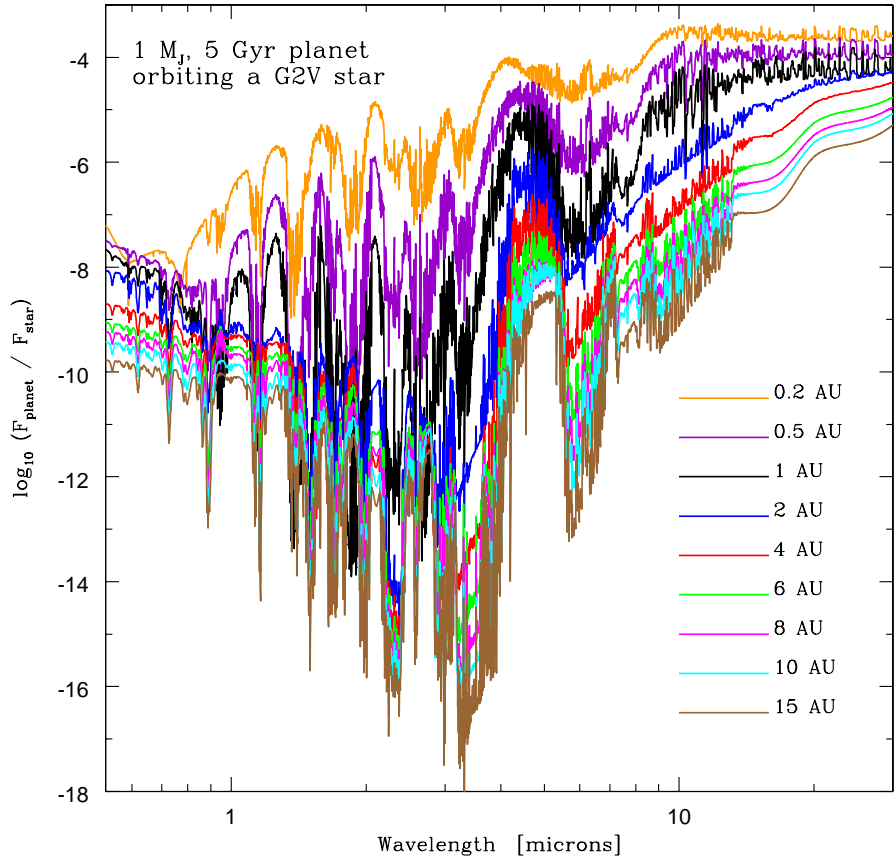}
 \caption{\small Theoretical planet to star flux ratios versus wavelength (in microns) from 0.5 \mic to 30 \mic
for a 1-\mj EGP with an age of 5 Gyr orbiting a G2V main sequence star similar to the Sun.
This figure portrays ratio spectra as a function of orbital diatance from 0.2 AU
to 15 AU.  Zero eccentricity is assumed and the planet spectra have been phase-averaged
as described in Sudarsky, Burrows, \& Hubeny (2003). The associated $T/P$ profiles are given in Fig. \ref{profiles}.
Table 1 in Burrows, Sudarsky, \& Hubeny (2004) lists the modal radii for the particles in the water and ammonia clouds.
Note that the planet/star flux ratio is most favorable in the
mid-infrared. [Taken from Burrows et al. (2004)]
\label{contrastd}}
 \end{figure*}

\begin{figure*}
 \epsscale{1.5}
 \plotone{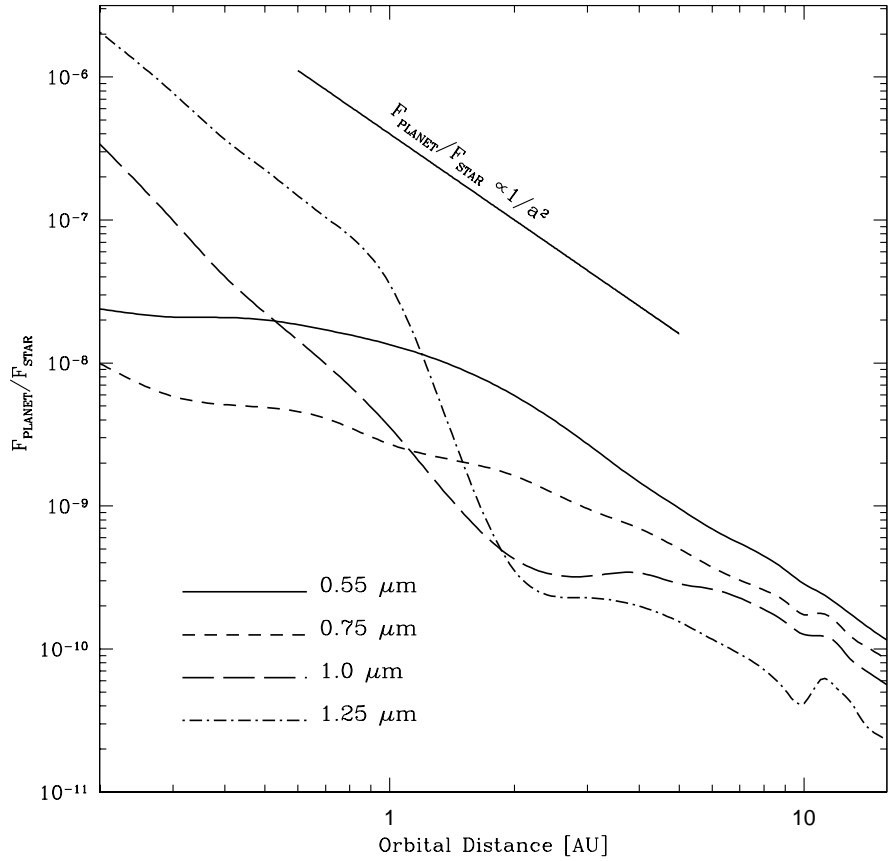}
 \caption{\small Planet/star flux ratio as a function of orbital distance at 0.55 $\mu$m, 0.75 $\mu$m,
1 $\mu$m, and 1.25 $\mu$m assuming a G2V central star.  In each case, the plotted
value corresponds to a planet at greatest elongation with an orbital inclination
of 80$^\circ$.  Note that the planet/star flux ratios do not follow a simple $1/a^2$ law.
[Taken from Sudarsky et al. (2005)]
\label{fig_ratiodist}}
 \end{figure*}

\begin{figure*}
\includegraphics[width=11.cm,angle=-90,clip=]{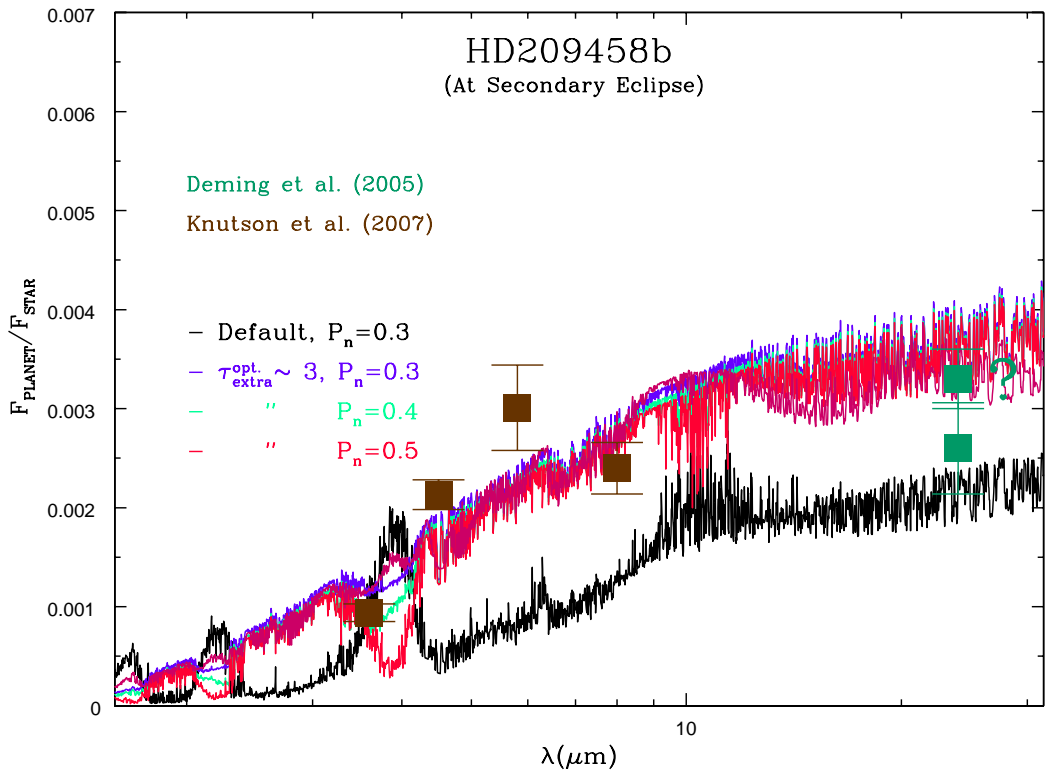}
\includegraphics[width=11.cm,angle=-90,clip=]{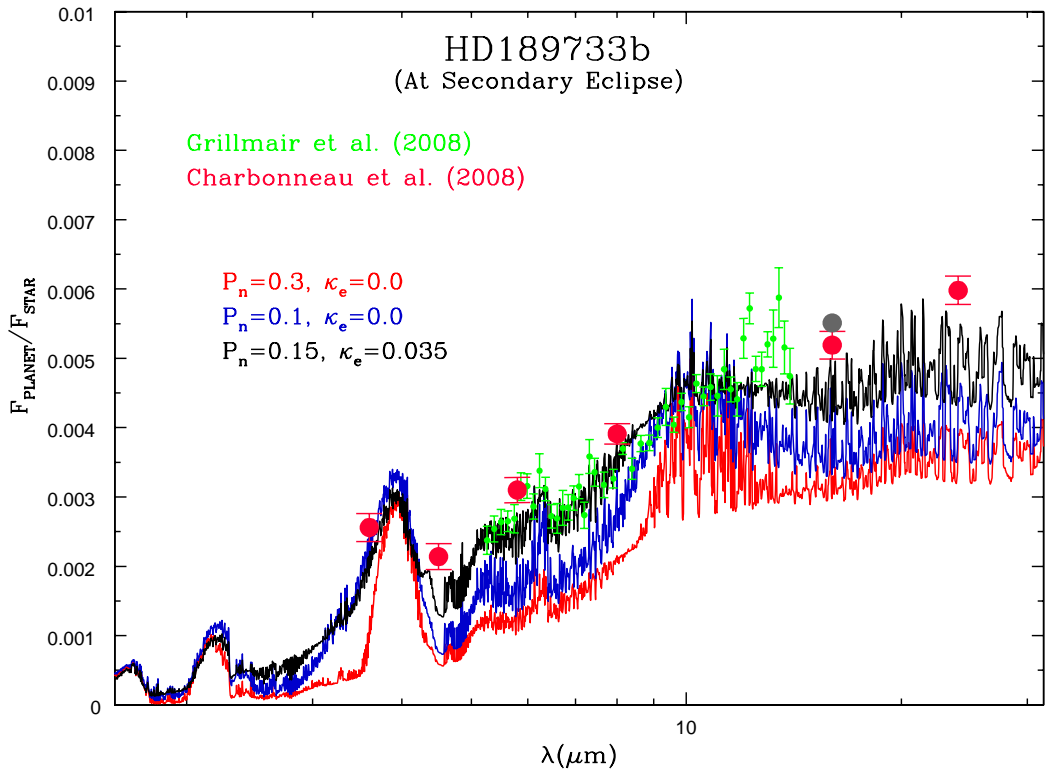}
\caption{\small
{\bf Top:} The planet-star flux ratios at secondary eclipse versus wavelength for
three models of the atmosphere of HD 209458b with inversions and for one model without an
extra upper-atmosphere absorber of any kind (lowest curve).
The three models with stratospheres have different values of P$_n$ (= 0.3,
0.4, and 0.5), but are otherwise the same.  The old, default model
has a P$_n$ of 0.3.  This figure demonstrates that models
with an extra upper-atmosphere absorber in the optical and with
P$_n \sgreat 0.35$ fit the data; the old model does not fit.
Superposed are the four IRAC points and the MIPS data at
24 $\mu$m from Deming et al. (2005) (square block on right). Also included, with a
question mark beside it, is a tentative update to this 24 $\mu$m flux point,
provided by Drake Deming (private communication).
{\bf Bottom:} Comparison of spectral observations with broadband
photometry and theoretical models of the dayside atmosphere of HD 189733b.
The black points show the mean (unweighted) flux ratios spectra from
$\sim$5 to $\sim$14 $\mu$m, taken by IRS on {\it Spitzer} and published in Grillmair
et al. (2008).  The plotted uncertainties reflect the
standard error in the mean in each wavelength bin. The filled red circles
show broadband measurements at 3.6, 4.5, 5.8, 8.0, 16, and 24 $\mu$m from
Charbonneau et al. (2008). Shown in color are atmospheric model
predictions for three values of a dayside-nightside heat redistribution
parameter, P$_n$, and two values for the extra upper-atmosphere
opacity, $\kappa_e$ (see Burrows, Budaj, \& Hubeny 2008 for an explanation).
See the text for discussion.
\label{hd189.209.spectra}}
\end{figure*}

\begin{figure*}
\includegraphics[width=6.5cm,angle=-90]{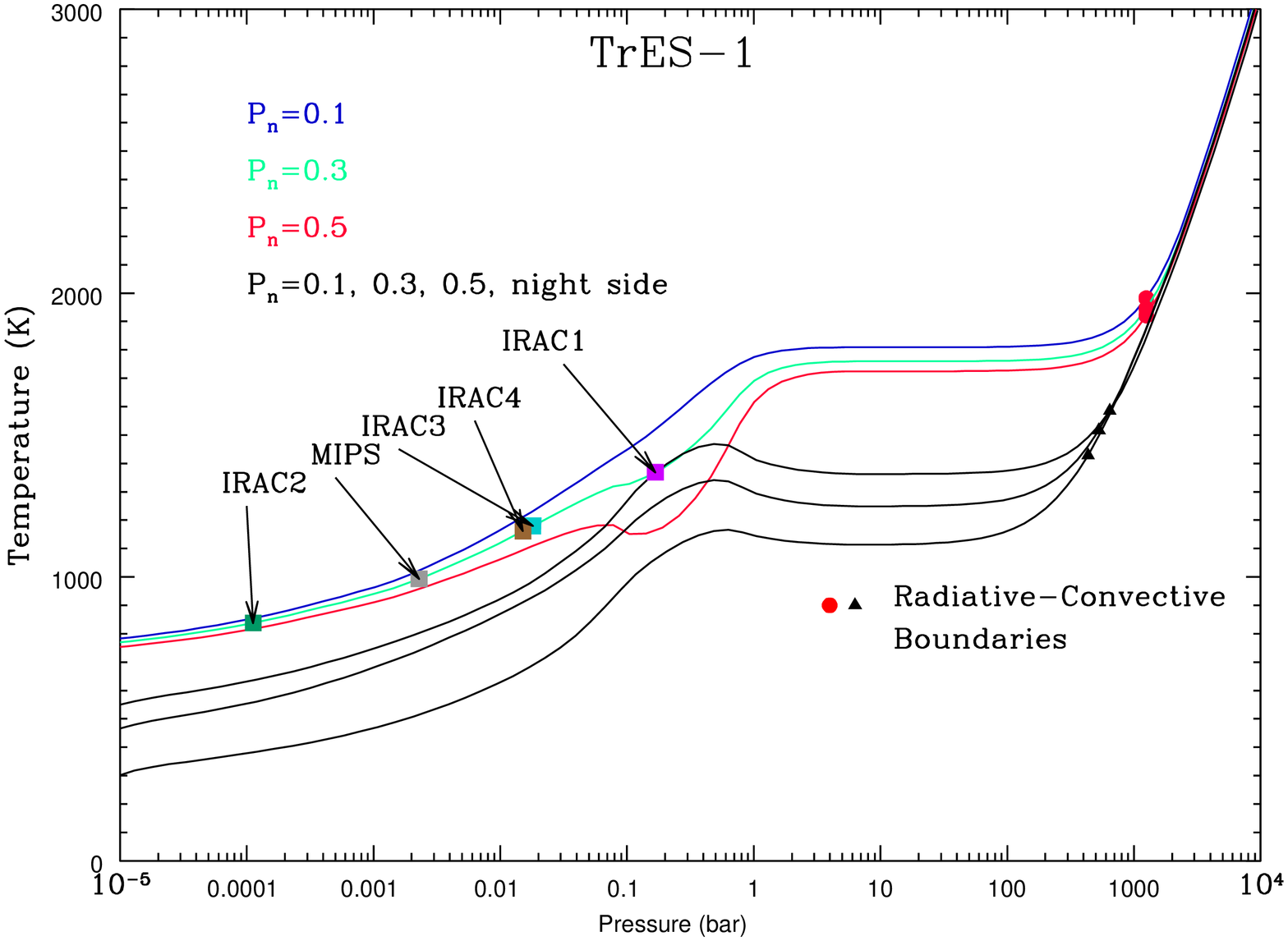}
\includegraphics[width=6.5cm,angle=-90]{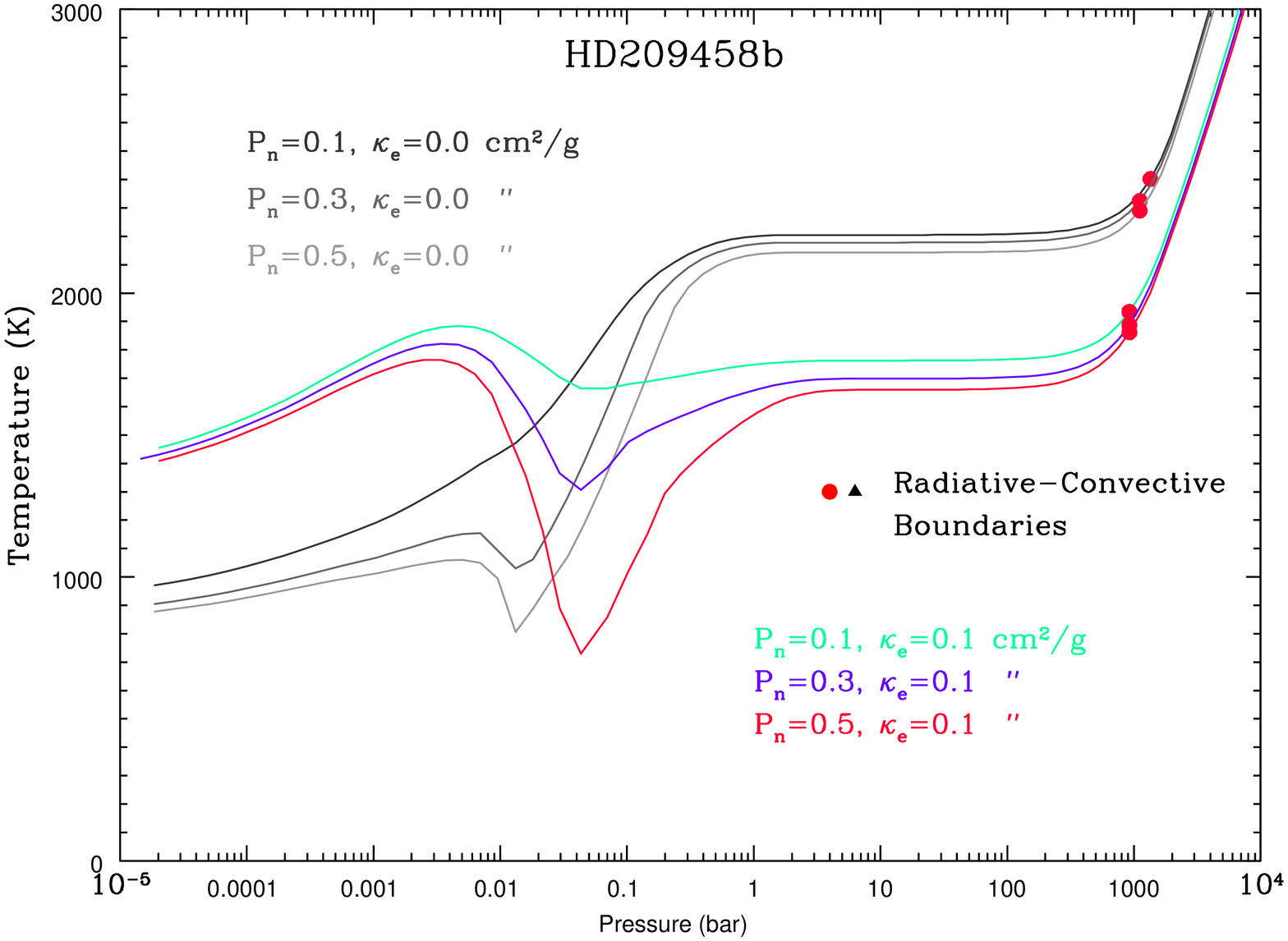}
\caption{\small Dayside (set of higher curves around 10 bars) 
and nightside (lower set of curves) temperature-pressure 
profiles for TrES-1 (left), as a representative of a transiting EGP 
without a thermal inversion at altitude and for HD 209458b (right), 
as a representative of a transiting EGP with such a thermal inversion.
These profiles incorporate the external substellar irradiation/flux and an internal
flux for the planet corresponding to the temperature of 75 K at
are included for HD 209458b. See text in for a discussion.  Figures 
taken from Burrows, Budaj, and Hubeny (2008), to which the reader is 
referred for further details.  
\label{tp.eps}}
\end{figure*}

\end{document}